\documentclass[12pt]{article}
\usepackage{amsmath} 
\usepackage{eufrak}
\usepackage{cite}
\usepackage{multicol}
\usepackage{a41}
\newcommand{\Li}{{\rm Li}}
\newcommand{\Sf}{{\rm S}_{1,2}}
\newcommand{\si}{{\rm sign}}

\newcommand{\bq}{\begin{equation}}
\newcommand{\eq}{\end{equation}}
\newcommand\beq{\begin{equation}}
\newcommand\eeq{\end{equation}}
\newcommand\bea{\begin{eqnarray}}
\newcommand\eea{\end{eqnarray}}

\newcommand\MV{\,\mbox{\bf M}}

\newcommand\SH{\,\mbox{$\sqcup \! \sqcup$}\,}
\include{epsfig}
\begin{document}
\setlength{\baselineskip}{0.515cm}
\sloppy
\thispagestyle{empty}
\begin{flushleft}
DESY 03-134 \hfill 
{\tt hep-ph/0311046}\\
October 2003\\
\end{flushleft}

\mbox{}
\vspace*{\fill}
\begin{center}
{\LARGE\bf Algebraic Relations Between Harmonic Sums }\\

\vspace{2mm}
{\LARGE\bf and Associated Quantities}

\vspace{4cm}
\large
Johannes Bl\"umlein

\vspace{1.5cm}
\normalsize
{\it  Deutsches Elektronen--Synchrotron, DESY,}\\
{\it  Platanenallee 6, D-15735 Zeuthen, Germany}

\end{center}
\normalsize
\vspace{\fill}
\begin{abstract}
\noindent
We derive the algebraic relations of alternating and non-alternating finite 
harmonic sums up to the sums of depth~6. All relations for the sums up to 
weight~6 are given in explicit form. These relations depend on the structure 
of the index sets of the harmonic sums only, but not on their value. They 
are therefore valid for all other mathematical objects which obey the same 
multiplication relation or can be obtained as a special case thereof, as the 
harmonic polylogarithms. We verify that the number of independent elements 
for a given index set can be determined by counting the Lyndon words which 
are associated to this set. The algebraic relations between the finite harmonic
sums can be used to reduce the high complexity of the expressions for the 
Mellin moments of the Wilson coefficients and splitting functions significantly 
for massless field theories as QED and QCD up to three loop and higher orders 
in the coupling constant and are also of importance for processes depending 
on more scales. The ratio of the number of independent sums thus obtained to 
the number of all sums for a given index set is found to be $\leq 1/d$ with 
$d$ the depth of the sum independently of the weight. The corresponding counting
relations are given in analytic form for all classes of harmonic sums to 
arbitrary depth and are tabulated up to depth $d=10$. 
\end{abstract}

\vspace{1mm}
\noindent

\vspace*{\fill}
\noindent
\numberwithin{equation}{section}
%%%%%%%%%%%%%%%%%%%%%%%%%%%%%%%%%%%%%%%%%%%%%%%%%%%%%%%%%%%%%%%%%%%%%%%
\newpage
%%%%%%%%%%%%%%%%%%%%%%%%%%%%%%%%%%%%%%%%%%%%%%%%%%%%%%%%%%%%%%%%%%%%%%%%
\section{Introduction}
%%%%%%%%%%%%%%%%%%%%%%%%%%%%%%%%%%%%%%%%%%%%%%%%%%%%%%%%%%%%%%%%%%%%%%%%
%

\vspace{1mm}
\noindent
Single scale problems in massless and massive perturbative calculations in
Quantum Field Theory can be expressed in terms of finite harmonic 
sums~\cite{HS1, HS2, HS3}. These sums occur in the $\epsilon$--expansion of
the integrals for higher order corrections to QCD splitting functions and
Wilson coefficients for space- and time-like processes~\cite{SF1} and the
amplitudes of important high energy 
scattering processes such as Bhabha scattering~\cite{BHA}, $pp \rightarrow 2$
jets~\cite{JET}, $pp \rightarrow \gamma \gamma$~\cite{PG}, 
Higgs production in hadron scattering~\cite{HIGGS},
light-by-light scattering~\cite{LLI}, and other QED processes~\cite{QEDP}. 
Multiple harmonic sums or  mathematical objects being related to them do 
generally emerge in Taylor expansions of higher transcendental functions 
occurring in loop integrals, see e.g.~\cite{HTF}. Unlike for 
representations in $x$--space multiple harmonic sums which are obtained 
after a {\sc Mellin} transform of the respective expressions account 
for the genuine {\sc Mellin} symmetry in massless field theory, the 
observation of which can lead to a considerable simplification of the 
these expressions. The multiple finite harmonic sums are defined by
%-------------------------------------------------------------------------
\begin{eqnarray}
\label{eqHS}
S_{a_1, \ldots, a_n}(N) = \sum_{k_1 = 1}^N \sum_{k_2 = 1}^{k_1} \ldots
\sum_{k_n = 1}^{k_{n-1}} \frac{\si(a_1)^{k_1}}{k_1^{|a_1|}} \ldots
\frac{\si(a_n)^{k_n}}{k_n^{|a_n|}}~.
\end{eqnarray}
%-------------------------------------------------------------------------
Here, $a_k$ are positive or negative integers and $N$ is a positive
even or odd integer depending on the observable under consideration. One
calls $n$ the {\sf depth} and $\sum_{k = 1}^n
|a_k|$ the {\sf weight} of a harmonic sum. Harmonic sums are associated to
{\sc Mellin} transforms of real functions or {\sc Schwartz}--distributions
$f(x)$~$\epsilon~{\cal S}'[0,1]$~\cite{DISTR}  
%-------------------------------------------------------------------------
\begin{eqnarray}
S_{a_1, \ldots, a_n}(N) = \int_0^1 dx~x^N~f_{a_1, \ldots, a_n}(x)
\end{eqnarray}
%-------------------------------------------------------------------------
which emerge in field theoretic calculations. Finite harmonic sums are
related to harmonic polylogarithms $H_{b_1, \ldots, b_n}(x)$~\cite{VR}.
Their $1/(1\pm x)$--weighted {\sc Mellin} transform yields harmonic sums.
The inverse {\sc Mellin} transform relates the harmonic sums to functions 
of {\sc Nielsen} integrals~\cite{NIELS}
and the variable $x$ at least for sums of weight $w \leq 4$ as shown in 
\cite{HS3}, and associated generalizations for higher weight.
{\sc Nielsen} integrals are a generalization of the usual
polylogarithms~\cite{POLYL}. In the limit $N \rightarrow \infty$ the
convergent multiple harmonic sums, i.e. those where $a_1 \neq 1$, yield
(multiple) Zeta--values $\zeta_{a_1, \ldots, a_n}$, which are also called 
{\sc  Euler--Zagier} sums~\cite{EZ}. A generalization of both
harmonic
polylogarithms and the {\sc Euler--Zagier} sums are the nested 
$Z$--sums~\cite{MUW}, which form a {\sc Hopf} algebra~\cite{HOPF,KAS}
and are related to Goncharov's multiple 
polylogarithms~\cite{GON}~\footnote{The two--dimensional harmonic polylogarithms 
\cite{GR} are a special case of the latter class of functions.}.
Likewise one may also consider polylogarithms of a different kind~\cite{LIS}.
\footnote{For a recent review see \cite{WALDS}.}

For many applications it is very convenient to express the result first in
analytic form in terms of finite harmonic sums or rational functions out of
them. This is even mandatory studying factorization--scheme invariant 
evolution~\cite{DLY}.
Harmonic sums may be analytically continued from the even or odd 
integers to complex
values of $N$~\cite{ANCONT} which, however, requires some effort in
general to obtain highly precise
representations. The resulting functions $f(x)$ can be easily obtained by
the inverse {\sc Mellin} transformation through a single numerical 
contour integral in
the complex $N$-plane. Since the number of harmonic sums up to weight and
depth $n$ is $3^{n}-1$ the number of necessary
analytic continuations of the respective
harmonic sums would grow rather rapidly. The finite harmonic sums obey
algebraic relations of different kind, which may be used to determine an
explicit (polynomial) representation of all sums of a given weight and
depth over a basis of independent sums. 

It is the aim of the present paper to calculate these representations for
all harmonic sums up to depth~6. The results will be applied to derive the
representations needed up to depth and weight~6 in explicit form leaving, 
however, the respective indices as general parameters. This level is 
expected to be sufficient to express the anomalous dimensions and massless 
coefficient 
functions, and similarly other quantities, up to $O(\alpha^3_i)$ in QED and 
QCD.       

The paper is organized as follows. In section~2 we derive the algebraic
relations of finite harmonic sums up to depth~6 in general form. There
are in principle two classes: the product--relations and those implied by 
integration by parts. We show that the latter relations are fully contained
in the former, since always all allowed index permutations can be considered. 
In section~3--6  the specific cases are considered for depth~3
to depth~6. Here we also complete foregoing investigations~\cite{HS3,ANCONT} 
concerning the analytic continuation of sums up to depth and weight~4. 
The relations obtained were extensively tested numerically. 
The relations for the finite harmonic sums may be directly applied to
obtain relations for the associated {\sc Euler--Zagier} sums in the
limit
$N \rightarrow \infty$.~\footnote{For positive $a_i$'s the finite
Zeta--values were calculated to weight~9 in \cite{BROAD}, to weight~10
in \cite{PET1} and to weight~12 in \cite{PET2}. Allowing general values 
of $a_i$ explicit relations were given up to weight~4 in \cite{GT,HS3}.
More recently also multiple Zeta--values with weights, which are $n$th
roots of unity were considered and the explicit results for the 
alternating case were given up to weight~7 in \cite{PET3}. All multiple 
Zeta--values up to weight~9 were obtained using {\tt FORM}~\cite{FORM},
see \cite{HS2}, and expressed in terms of the more familiar numbers 
(symbols) $S_1(\infty), \ln(2), \zeta(2), \zeta(3), \Li_4(1/2), \zeta(5),
\Li_5(1/2), \Li_6(1/2), S_{-5,-1}(\infty), \zeta(7), \Li_7(1/2),
S_{-5,1,1}(\infty)$ and $S_{5,-1,-1}(\infty)$ as also in \cite{GT,HS3} 
and
complementary to \cite{PET2,PET3}. For a quantum field theoretic
representation of {\sc Euler-Zagier} sums see \cite{MUSC}.}
Moreover, as all the relations considered in the 
present paper derive from the structure of the respective index set and do 
not depend on the specific value of the harmonic sums these 
relations 
hold for {\sf all} other mathematical objects which obey the same 
multiplication relation (\ref{eqPROD}). The main product--relation considered 
in the present paper (\ref{eqPROD}) reduces to a simpler one
(\ref{eqHPR}), which is valid for harmonic polylogarithms~\cite{VR} by 
deleting all terms which contain the
$\wedge$--operator (\ref{eqWEDG}) 
in the index set. Therefore all algebraic relations 
derived cover the respective algebraic relations for harmonic polylogarithms
(\ref{eqHPR}).
Similarly, the algebraic relations for $Z$--sums with $x_i = 1, \forall~i$ 
are easily derived, e.g. using Eq.~(6) of Ref.~\cite{MUW}.
In section~7 we briefly show that the 
number of independent harmonic sums for a given index set can be counted
determining the number of {\sc Lyndon} words~\cite{LYND1,LYND2} which 
belong to this 
set. Section~8 contains the conclusions.  
In appendix~A we summarize the numbers of sums and basic sums for all
individual index pattern from depth $d=7$ to 10.
Appendix~B contains an overview on all specific harmonic 
sums up to depth and weight~6.

%%%%%%%%%%%%%%%%%%%%%%%%%%%%%%%%%%%%%%%%%%%%%%%%%%%%%%%%%%%%%%%%%%%%%%%%
\section{General Algebraic Relations}
%%%%%%%%%%%%%%%%%%%%%%%%%%%%%%%%%%%%%%%%%%%%%%%%%%%%%%%%%%%%%%%%%%%%%%%%
%
%%%%%%%%%%%%%%%%%%%%%%%%%%%%%%%%%%%%%%%%%%%%%%%%%%%%%%%%%%%%%%%%%%%%%%%%
\subsection{Product--Relations}
%%%%%%%%%%%%%%%%%%%%%%%%%%%%%%%%%%%%%%%%%%%%%%%%%%%%%%%%%%%%%%%%%%%%%%%%

\vspace{1mm}
\noindent
%------------------------------------------------------------------------
The product of two finite harmonic sums (\ref{eqHS}) yields
%------------------------------------------------------------------------
\begin{eqnarray}
\label{eqPROD}
S_{a_1, \ldots, a_n}(N) \cdot S_{b_1, \ldots, b_m}(N)
&=& \sum_{l_1=1}^N \frac{\si(a_1)^{l_1}}{l_1^{|a_1|}}
  S_{a_2, \ldots, a_n}(l_1)\, S_{b_1, \ldots, b_m}(l_1) \nonumber\\ & &
+ \sum_{l_2=1}^N \frac{\si(b_1)^{l_2}}{l_2^{|b_1|}}  
  S_{a_1, \ldots, a_n}(l_2)\, S_{b_2, \ldots, b_m}(l_2) \nonumber\\ & &
- \sum_{l=1}^N \frac{[\si(a_1) \si(b_1)]^l}{l^{|a_1|+|b_1|}}  
  S_{a_2, \ldots, a_n}(l)\, S_{b_2, \ldots, b_m}(l)~.
\end{eqnarray}
%------------------------------------------------------------------------
One proves by induction that the r.h.s. of Eq.~(\ref{eqPROD}) 
consists out of a linear combination of harmonic sums of the argument $N$
and depth $m+n$ or lower. The prerequisite is obtained choosing $n=1$~,
%------------------------------------------------------------------------
\begin{eqnarray}
\label{eqPROD1}
S_{a_1}(N) \cdot S_{b_1, \ldots, b_m}(N)
&=& S_{a_1, b_1, \ldots, b_m}(N) + S_{b_1, a_1, b_2, \ldots, b_m}(N)
+ \ldots + S_{b_1, b_2,  \ldots, b_m, a_1}(N) \nonumber\\
& & -S_{a_1 \wedge b_1, b_2, \ldots, b_m}(N) - \ldots - 
S_{b_1, b_2, \ldots, a_1 \wedge b_m}(N)~.
\end{eqnarray}
%------------------------------------------------------------------------
Here the symbol $\wedge$ is defined as
%------------------------------------------------------------------------
\begin{eqnarray}
\label{eqWEDG}
a \wedge b = \si(a) \si(b) \left(|a| + |b|\right)~.
\end{eqnarray}
%------------------------------------------------------------------------
Relations similar to (\ref{eqPROD1}) were investigated by {\sc Fa\`a di
Bruno}~\cite{FDB} for the roots of algebraic equations~\footnote{For
related work on the invariant theory of the algebraic equations with
various variables see \cite{CLGOR}.}
and {\sc Hoffman}~\cite{HOF} for {\sc Euler--Zagier} sums.
Eq.~(\ref{eqPROD1}) may be
used to establish that the linear combination of all permutations over a
given index set of finite harmonic sums can be represented in terms of a 
polynomial of {\sf single} harmonic sums~\cite{HS3, PWILL}\footnote{In
\cite{PWILL} the complete permutations were derived for arbitrary indices 
up to depth $n=10$. The representation for 
$n=2$ was first derived in \cite{EUL}, that for
$n=3$ in \cite{IND} and for $n=4$ in \cite{HS3}.}, cf.~section~{2.3}. This holds
likewise also for positive integer sums~\cite{RAMAN}. If all the indices 
of a finite harmonic sum are equal, it obeys a
determinant--representation, see \cite{HS3}. Another representation was
given in \cite{HS2}.

We introduce the {\sf shuffle product} $\SH$ of a single  and a 
general finite harmonic 
sum 
%------------------------------------------------------------------------
\begin{eqnarray}
\label{eqSTAF}
S_{a_1}(N) \SH S_{b_1, \ldots, b_m}(N)
&=& S_{a_1, b_1, \ldots, b_m}(N) + S_{b_1, a_1, b_2, \ldots, b_m}(N)
+ \ldots + S_{b_1, b_2,  \ldots, b_m, a_1}(N)
\end{eqnarray}
%------------------------------------------------------------------------
which is the linear combination of the sums of depth $m+1$ which are 
generated by Eq.~(\ref{eqPROD1}). The shuffle product of two harmonic sums
of depth $n$ and $m$, $S_{a_1, \ldots, a_n}(N)$ and $S_{b_1, \ldots,
b_m}(N)$, is
the sum of all harmonic sums of depth $m+n$ in the index set of which $a_i$
occurs left of $a_j$ for $i < j$ and likewise for 
$b_k$ and $b_l$ for $k < l$.~\footnote{Shuffle 
products were introduced in \cite{SHUFF}, 
see also \cite{LYND1}. For the application of shuffles for multiple zeta
values see e.g. \cite{BBBL}.}
Shuffle products are symmetric
%------------------------------------------------------------------------
\begin{eqnarray}
\label{eqCOMU}
S_{a_1, \ldots, a_n}(N) \SH S_{b_1, \ldots, b_m}(N)
= S_{b_1, \ldots, b_m}(N) \SH S_{a_1, \ldots, a_n}(N)~,   
\end{eqnarray}
%------------------------------------------------------------------------
i.e. the commutation relation holds. The set of harmonic
sums is extended by the constant {\sf 1}, the empty sum, which forms the 
unit element
%------------------------------------------------------------------------
\begin{eqnarray}
\label{eqUNI}
{\sf 1}~\SH~S_{a_1, \ldots, a_n}(N)
= S_{a_1, \ldots, a_n}~.
\end{eqnarray}
%------------------------------------------------------------------------

We consider the product of the double sum $S_{a_1, a_2}(N)$ with a
general harmonic sum. The first term of the r.h.s. of Eq.~(\ref{eqPROD}) reduces
to a sum
over
(\ref{eqPROD1}) which is a linear combination of harmonic sums, likewise the
third term. Inverting the relation for the second term of the r.h.s. of
(\ref{eqPROD1}) one may perform the complete recursion for this case into a
linear combination of harmonic sums as well. Because products  
$S_{a_1, \ldots a_k}(N) \cdot S_{b_1, \ldots, b_l}(N)$ with either $k < n$ or
$l < m$ can be represented in terms of linear combinations of harmonic sums,
Eq.~(\ref{eqPROD1}) is also a linear combination of harmonic sums. 

%-------------------------------------------

The products of depth~$n$ induce algebraic relations between the single harmonic
sums of depth $n$, which allow their calculation in terms of sums of the same and
lower depth. Associated to this the corresponding shuffle products of 
harmonic sums occur, which are

%\vspace{1mm} \noindent
%{\sf Depth~2:}\\
%------------------------------------------------------------------------------$
\begin{eqnarray}
{\sf Depth~2:} \hspace{3cm}& &  \nonumber\\
S_{a_1}(N) \SH S_{a_2}(N) &=& S_{a_1, a_2}(N) + S_{a_2, a_1}(N) 
\end{eqnarray}
\begin{eqnarray}
{\sf Depth~3:} \hspace{3cm}& &  \nonumber\\
\label{eqD3}
S_{a_1}(N) \SH S_{a_2, a_3}(N) &=& S_{a_1, a_2, a_3}(N) + S_{a_2, a_1, 
a_3}(N) + S_{a_2, a_3, a_1}(N)
\\
%\end{eqnarray}
%------------------------------------------------------------------------------
%\begin{eqnarray}
{\sf Depth~4:} \hspace{3cm}& &  \nonumber\\
\label{eqD4A}
S_{a_1}(N) \SH S_{a_2, a_3, a_4}(N) &=& S_{a_1, a_2, a_3, a_4}(N) + 
S_{a_2, a_1, a_3, a_4}(N) + S_{a_2, a_3, a_1, a_4}(N) +  S_{a_2, a_3, 
a_4, a_1}(N)\nonumber\\
\\
\label{eqD4B}
S_{a_1, a_2}(N) \SH S_{a_3, a_4}(N) &=& S_{a_1, a_2, a_3, a_4}(N)
+ S_{a_1, a_3, a_2, a_4}(N) + S_{a_1, a_3, a_4, a_2}(N)\nonumber\\ &+&
S_{a_3, a_4, a_1, a_2}(N) + S_{a_3, a_1, a_4, a_2}(N) +
S_{a_3, a_1, a_2, a_4}(N)%\\
%\nonumber\\
%\end{eqnarray}
%\begin{eqnarray}
\\
{\sf Depth~5:} \hspace{3cm}& &  \nonumber\\   
\label{eqD5A}
S_{a_1}(N) \SH S_{a_2, a_3, a_4, a_5}(N) &=& S_{a_1, a_2, a_3, a_4, a_5}(N) +
S_{a_2, a_1, a_3, a_4, a_5}(N) + S_{a_2, a_3, a_1, a_4, a_5}(N)
\nonumber\\ &+& 
S_{a_2, a_3, a_4, a_1, a_5}(N) +  S_{a_2, a_3, a_4, a_5, a_1}(N)
%\nonumber\\
\\
\label{eqD5B}
S_{a_1, a_2}(N) \SH S_{a_3, a_4, a_5}(N) &=&
S_{a_1, a_2, a_3, a_4, a_5}(N) +
S_{a_1, a_3, a_2, a_4, a_5}(N) +
S_{a_1, a_3, a_4, a_2, a_5}(N)\nonumber\\ &+& 
S_{a_1, a_3, a_4, a_5, a_2}(N) +
S_{a_3, a_1, a_2, a_4, a_5}(N) +
S_{a_3, a_1, a_4, a_2, a_5}(N) \nonumber\\ &+&
S_{a_3, a_1, a_4, a_5, a_2}(N) +
S_{a_3, a_4, a_5, a_1, a_2}(N) +
S_{a_3, a_4, a_1, a_5, a_2}(N) \nonumber\\ &+&
S_{a_3, a_4, a_1, a_2, a_5}(N) \\
{\sf Depth~6:} \hspace{3cm}& &  \nonumber\\
\label{eqD6A}
S_{a_1}(N) \SH S_{a_2, a_3, a_4, a_5, a_6}(N) &=& 
S_{a_1, a_2, a_3, a_4, a_5, a_6}(N) +
S_{a_2, a_1, a_3, a_4, a_5, a_6}(N) + S_{a_2, a_3, a_1, a_4, a_5, a_6}(N)
\nonumber\\ &+&
S_{a_2, a_3, a_4, a_1, a_5, a_6}(N) +  S_{a_2, a_3, a_4, a_5, a_1, a_6}(N)
+  S_{a_2, a_3, a_4, a_5, a_6, a_1}(N)
\nonumber\\
\\
%\end{eqnarray}
%\begin{eqnarray}
\label{eqD6B}
S_{a_1, a_2}(N) \SH S_{a_3, a_4, a_5, a_6}(N) &=& 
S_{a_1, a_2, a_3, a_4, a_5, a_6}(N)
       +S_{a_1, a_3, a_2, a_4, a_5, a_6}(N)
       +S_{a_1, a_3, a_4, a_2, a_5, a_6}(N) \nonumber\\ &+&
        S_{a_1, a_3, a_4, a_5, a_2, a_6}(N)
       +S_{a_1, a_3, a_4, a_5, a_6, a_2}(N)
       +S_{a_3, a_1, a_2, a_4, a_5, a_6}(N) \nonumber\\ &+&
        S_{a_3, a_1, a_4, a_2, a_5, a_6}(N)
       +S_{a_3, a_1, a_4, a_5, a_2, a_6}(N)
       +S_{a_3, a_1, a_4, a_5, a_6, a_2}(N) \nonumber\\ &+&
        S_{a_3, a_4, a_1, a_2, a_5, a_6}(N)
       +S_{a_3, a_4, a_1, a_5, a_2, a_6}(N)
       +S_{a_3, a_4, a_1, a_5, a_6, a_2}(N) \nonumber\\ &+&
        S_{a_3, a_4, a_5, a_6, a_1, a_2}(N)
       +S_{a_3, a_4, a_5, a_1, a_6, a_2}(N)
       +S_{a_3, a_4, a_5, a_1, a_2, a_6}(N)
\nonumber\\
\\
%\end{eqnarray}
%\begin{eqnarray}
\label{eqD6C}
S_{a_1, a_2, a_3}(N) \SH S_{a_4, a_5, a_6}(N) &=& 
S_{a_1, a_2, a_3, a_4, a_5, a_6}(N)
     +S_{a_1, a_2, a_4, a_3, a_5, a_6}(N)
     +S_{a_1, a_2, a_4, a_5, a_3, a_6}(N) \nonumber\\ &+&
      S_{a_1, a_2, a_4, a_5, a_6, a_3}(N)
     +S_{a_1, a_4, a_2, a_3, a_5, a_6}(N)
     +S_{a_1, a_4, a_2, a_5, a_3, a_6}(N) \nonumber\\
%\end{eqnarray}
%\begin{eqnarray}
&+&
      S_{a_1, a_4, a_2, a_5, a_6, a_3}(N)
     +S_{a_1, a_4, a_5, a_6, a_2, a_3}(N)
     +S_{a_1, a_4, a_5, a_2, a_6, a_3}(N) \nonumber\\ &+&
      S_{a_1, a_4, a_5, a_2, a_3, a_6}(N)
     +S_{a_4, a_5, a_6, a_1, a_2, a_3}(N)
     +S_{a_4, a_5, a_1, a_6, a_2, a_3}(N) \nonumber\\ &+&
      S_{a_4, a_5, a_1, a_2, a_6, a_3}(N)
     +S_{a_4, a_5, a_1, a_2, a_3, a_6}(N)
     +S_{a_4, a_1, a_5, a_6, a_2, a_3}(N) \nonumber\\ &+&
      S_{a_4, a_1, a_5, a_2, a_6, a_3}(N)
     +S_{a_4, a_1, a_5, a_2, a_3, a_6}(N)
     +S_{a_4, a_1, a_2, a_3, a_5, a_6}(N) \nonumber\\ &+&
      S_{a_4, a_1, a_2, a_5, a_3, a_6}(N)
     +S_{a_4, a_1, a_2, a_5, a_6, a_3}(N)~.
\end{eqnarray}
%------------------------------------------------------------------------------$ 
The number of harmonic sums $n{\{m_1 \SH m_2\}}$ occurring in the r.h.s. 
of 
the binary shuffle 
products of a sum of depth $m_1$ and $m_2$ are 
%------------------------------------------------------------------------------$
\begin{eqnarray}
n{\{m_1 \SH m_2\}} = \binom{m_1 + m_2}{m_1}~.
\end{eqnarray}
%------------------------------------------------------------------------------

For a given depth $n$ one may consider also shuffle products of more than 
two factors. For $n=3$ the triple shuffle product of single sums induces 
the sum over the complete permutation of the three indices. This 
combination is linearly dependent of either two equations of type 
(\ref{eqD3}). More generally, the shuffle  product of $m$ single sums 
leads to the sum of all harmonic sums of depth $m$ permuting the whole 
index set. As we will later consider entire sets of all index permutations of 
the relations (\ref{eqD3}--\ref{eqD6C}) this combination is always 
linearly dependent and does not lead to a new relation. For $n=4$ the
product $S_{a_1}(N) \SH S_{a_2}(N) \SH S_{a_3, a_4}(N)$ is related to the 
combination $S_{a_1, a_2}(N) \SH S_{a_3, a_4}(N) + S_{a_2, a_1}(N) \SH 
S_{a_3, a_4}(N)$ and forms therefore a linear combination out of the 
permutation set for (\ref{eqD4B}). Likewise the same argument holds for
(\ref{eqD5B}, \ref{eqD6B}) for $n = 5,6$, respectively. For $n=5$ the 
product $S_{a_1}(N) \SH S_{a_2}(N) \SH S_{a_3}(N) \SH 
S_{a_4, a_5}(N)$ is linearly dependent of the permutations generated
by (\ref{eqD5B}) and similarly for the partial products of triple single 
sums for $n=6$ being linearly dependent of the permutations of 
(\ref{eqD6C}). Up to depth 6 only the binary shuffles are found to 
contribute. 

In the following we list the genuine relations of the finite harmonic 
sums up to depth 6. Sums of this type emerge in massless 3--loop 
calculations.
%----------------------------------------------------------------------------------
\begin{eqnarray}
{\sf Depth~2:} \hspace{3cm} && \nonumber\\
S_{a_1}(N) \SH S_{a_2}(N) &-& S_{a_1}(N) S_{a_2}(N) - S_{a_1 \wedge a_2}(N)
= 0~~~~~~\mbox{\cite{EUL}} \\
{\sf Depth~3:} \hspace{3cm} && \nonumber\\
S_{a_1}(N) \SH  S_{a_2, a_3}(N) &-&
S_{a_1}(N) S_{a_2, a_3}(N) - S_{a_1 \wedge a_2, a_3}(N) 
- S_{a_2, a_1 \wedge a_3}(N) = 0 \\
{\sf Depth~4:} \hspace{3cm} && \nonumber\\
\label{eqR4A}
S_{a_1}(N) \SH S_{a_2, a_3, a_4}(N) &-&
S_{a_1}(N) S_{a_2, a_3, a_4}(N) 
- S_{a_1 \wedge a_2, a_3, a_4}(N)
-S_{a_2, a_1 \wedge a_3, a_4}(N) \nonumber\\ &-&
S_{a_2, a_3, a_1 \wedge a_4}(N) 
=0 \\
\label{eqR4B}
S_{a_1, a_2}(N) \SH S_{a_3, a_4}(N) &-&
S_{a_1, a_2}(N) S_{a_3, a_4}(N) - S_{a_1, a_2 \wedge a_3, a_4}(N)
- S_{a_1, a_3, a_2 \wedge a_4}(N) \nonumber\\
&-& S_{a_3, a_1 \wedge a_4, a_2}(N) - S_{a_3, a_1, a_2 \wedge a_4}(N)
- S_{a_1 \wedge a_3, a_2, a_4}(N) \nonumber\\
&-& S_{a_1 \wedge a_3, a_4, a_2}(N) + S_{a_1 \wedge a_3, a_2 \wedge a_4}
  =0    \\
{\sf Depth~5:} \hspace{3cm} && \nonumber\\
S_{a_1}(N) \SH S_{a_2, a_3, a_4, a_5}(N) &-&
S_{a_1}(N) S_{a_2, a_3, a_4, a_5}(N)
- S_{a_1 \wedge a_2, a_3, a_4, a_5}(N)
-S_{a_2, a_1 \wedge a_3, a_4, a_5}(N) \nonumber\\ &-&
S_{a_2, a_3, a_1 \wedge a_4, a_5}(N)
- S_{a_2, a_3, a_4, a_1 \wedge a_5}(N)
=0 \\
%\end{eqnarray}
%\begin{eqnarray}
S_{a_1, a_2}(N) \SH S_{a_3, a_4, a_5}(N) &-&
S_{a_1, a_2 \wedge a_3, a_4, a_5}(N) - S_{a_1, a_3, a_2 \wedge a_4, 
a_5}(N)
- S_{a_1, a_3, a_4, a_2 \wedge a_5}(N) \nonumber\\
&-& S_{a_3, a_1, a_2 \wedge a_4, a_5}(N) - S_{a_3, a_1, a_4, a_2 \wedge 
a_5}(N) -S_{a_3, a_4, a_1 \wedge a_5, a_2}(N) \nonumber\\
%\end{eqnarray}
%\begin{eqnarray}
&-& S_{a_3, a_4, a_1, a_2 \wedge a_5}(N) - S_{a_3, a_1 \wedge a_4, a_2, 
a_5} (N) - S_{a_3, a_1 \wedge a_4, a_5 , a_2}(N) \nonumber\\
&-& S_{a_1, a_2}(N) S_{a_3, a_4, a_5}(N) - S_{a_1 \wedge a_3, a_2, a_4, 
a_5}(N)-S_{a_1 \wedge a_3, a_4, a_2, a_5}(N) \nonumber\\ &-&
S_{a_1 \wedge a_3, a_4, a_5, a_2}(N)
+S_{a_1 \wedge a_3, a_2 \wedge a_4, a_5}(N) + S_{a_1 \wedge a_3, a_4, 
a_2 \wedge a_5}(N)
 = 0    \nonumber\\
\\
%\end{eqnarray}  
%\begin{eqnarray}
%
%------------------------------------------------------------------
%
{\sf Depth~6:} \hspace{3cm} && \nonumber\\      
S_{a_1}(N) \SH S_{a_2, a_3, a_4, a_5, a_6}(N) &-& 
S_{a_1}(N) S_{a_2, a_3, a_4, a_5, a_6}(N) - S_{a_1 \wedge a_2, a_3, a_4, 
a_5, a_6}(N) \nonumber\\ &-&
 S_{a_2, a_1 \wedge a_3, a_4, a_5, a_6}(N) -
S_{a_2, a_3, a_1 \wedge a_4, a_5, a_6}(N) \nonumber\\ &-&
S_{a_2, a_3, a_4, a_1 \wedge a_5, a_6}(N) 
-S_{a_2, a_3, a_4, a_5, a_1 \wedge a_6}(N)
= 0  
\\
%\end{eqnarray}
%\begin{eqnarray}
%
%-----------
%
S_{a_1, a_2}(N) \SH S_{a_3, a_4, a_5, a_6}(N) &-& 
        S_{a_1, a_2 \wedge a_3, a_4, a_5, a_6}(N)
       -S_{a_1, a_3, a_2 \wedge a_4, a_5, a_6}(N)
       -S_{a_1, a_3, a_4, a_2 \wedge a_5, a_6}(N) \nonumber\\ &-&
        S_{a_1, a_3, a_4, a_5, a_2 \wedge a_6}(N)
       -S_{a_3, a_1, a_2 \wedge a_4, a_5, a_6}(N)
       -S_{a_3, a_1, a_4, a_2 \wedge a_5, a_6}(N) \nonumber\\ &-&
        S_{a_3, a_1, a_4, a_5, a_2 \wedge a_6}(N)
       -S_{a_3, a_4, a_1, a_2 \wedge a_5, a_6}(N)
       -S_{a_3, a_4, a_1, a_5, a_2 \wedge a_6}(N) \nonumber\\ &-&
        S_{a_3, a_4, a_5, a_1 \wedge a_6, a_2}(N)
       -S_{a_3, a_4, a_5, a_1, a_2 \wedge a_6}(N)
       -S_{a_3, a_4, a_1 \wedge a_5, a_2, a_6}(N) \nonumber\\ &-&
        S_{a_3, a_4, a_1 \wedge a_5, a_6, a_2}(N)
       -S_{a_3, a_1 \wedge a_4, a_2, a_5, a_6}(N)
       -S_{a_3, a_1 \wedge a_4, a_5, a_2, a_6}(N) \nonumber\\
        &-&
        S_{a_3, a_1 \wedge a_4, a_5, a_6, a_2}(N)
       -S_{a_1 \wedge a_3, a_2, a_4, a_5, a_6}(N)
       -S_{a_1 \wedge a_3, a_4, a_2, a_5, a_6}(N) \nonumber%\\ 
\end{eqnarray}
\begin{eqnarray}
        &-&
        S_{a_1 \wedge a_3, a_4, a_5, a_2, a_6}(N)
       -S_{a_1 \wedge a_3, a_4, a_5, a_6, a_2}(N) 
       +S_{a_3, a_4, a_1 \wedge a_5, a_2 \wedge a_6}(N) \nonumber\\ &+&
        S_{a_3, a_1 \wedge a_4, a_2 \wedge a_5, a_6}(N)
       +S_{a_3, a_1 \wedge a_4, a_5, a_2 \wedge a_6}(N) 
       +S_{a_3, a_4, a_1 \wedge a_5, a_2 \wedge a_6}(N) \nonumber\\ &+&
        S_{a_1 \wedge a_3, a_2 \wedge a_4, a_5, a_6}(N)
       +S_{a_1 \wedge a_3, a_4, a_2 \wedge a_5, a_6}(N)
       +S_{a_1 \wedge a_3, a_4, a_5, a_2 \wedge a_6}(N) \nonumber\\ &-&
        S_{a_1, a_2}(N) S_{a_3, a_4, a_5, a_6}(N) = 0  \\
%\end{eqnarray}
%\begin{eqnarray}
%
%-----------
%
S_{a_1, a_2, a_3}(N) \SH S_{a_4, a_5, a_6}(N) &-& 
      S_{a_1, a_2, a_3 \wedge a_4, a_5, a_6}(N)  
     -S_{a_1, a_2, a_4, a_3 \wedge a_5, a_6}(N)  
     -S_{a_1, a_2, a_4, a_5,a_3 \wedge  a_6}(N) \nonumber\\ &-&  
      S_{a_1, a_4, a_2, a_3 \wedge a_5, a_6}(N)  
     -S_{a_1, a_4, a_2, a_5,a_3 \wedge  a_6}(N)  
     -S_{a_1, a_4, a_5, a_2 \wedge a_6, a_3}(N) \nonumber\\ &-&
      S_{a_1, a_4, a_5, a_2, a_3 \wedge a_6}(N)
     -S_{a_1, a_4, a_2 \wedge a_5, a_3, a_6}(N)
     -S_{a_1, a_4, a_2 \wedge a_5, a_6, a_3}(N) \nonumber\\ &-&
      S_{a_1, a_2 \wedge a_4, a_3, a_5, a_6}(N)
     -S_{a_1, a_2 \wedge a_4, a_5, a_3, a_6}(N)
     -S_{a_1, a_2 \wedge a_4, a_5, a_6, a_3}(N) \nonumber\\ &-&
      S_{a_4, a_5, a_1 \wedge a_6, a_2, a_3}(N)
     -S_{a_4, a_5, a_1, a_2 \wedge a_6, a_3}(N)
     -S_{a_4, a_5, a_1, a_2, a_3 \wedge a_6}(N) \nonumber\\ &-&
      S_{a_4, a_1, a_5, a_2 \wedge a_6, a_3}(N)
     -S_{a_4, a_1, a_5, a_2,a_3 \wedge  a_6}(N)
     -S_{a_4, a_1, a_2, a_3 \wedge a_5, a_6}(N) \nonumber\\ &-&
      S_{a_4, a_1, a_2, a_5, a_3 \wedge a_6}(N)
     -S_{a_4, a_1, a_2 \wedge a_5, a_6, a_3}(N)
     -S_{a_4, a_1, a_2 \wedge a_5, a_3, a_6}(N) \nonumber\\ &-&
      S_{a_4, a_1 \wedge a_5, a_6, a_2, a_3}(N)
     -S_{a_4, a_1 \wedge a_5, a_2, a_6, a_3}(N)
     -S_{a_4, a_1 \wedge a_5, a_2, a_3, a_6}(N) \nonumber\\ &-&
      S_{a_1 \wedge a_4, a_2, a_3, a_5, a_6}(N)
     -S_{a_1 \wedge a_4, a_2, a_5, a_3, a_6}(N)
     -S_{a_1 \wedge a_4, a_2, a_5, a_6, a_3}(N) \nonumber\\ &-&
      S_{a_1 \wedge a_4, a_5, a_6, a_2, a_3}(N)
     -S_{a_1 \wedge a_4, a_5, a_2, a_6, a_3}(N) 
     -S_{a_1 \wedge a_4, a_5, a_2, a_3, a_6}(N)       \nonumber\\ &+&
      S_{a_1, a_4, a_2 \wedge a_5, a_3 \wedge a_6}(N)
     +S_{a_1, a_2 \wedge a_4, a_3 \wedge a_5, a_6}(N)
     +S_{a_1, a_2 \wedge a_4, a_5, a_3 \wedge a_6}(N) \nonumber\\ &+&
      S_{a_4, a_1, a_2 \wedge a_5, a_3 \wedge a_6}(N)
     +S_{a_4, a_1 \wedge a_5, a_2 \wedge a_6, a_3}(N)
     +S_{a_4, a_1 \wedge a_5, a_2, a_3 \wedge a_6}(N) \nonumber\\ &+&
      S_{a_1 \wedge a_4, a_2, a_3 \wedge a_5, a_6}(N)
     +S_{a_1 \wedge a_4, a_2, a_5, a_3 \wedge a_6}(N)
     +S_{a_1 \wedge a_4, a_5, a_2 \wedge a_6, a_3}(N) \nonumber\\ &+&
      S_{a_1 \wedge a_4, a_5, a_2, a_3 \wedge a_6}(N)
     +S_{a_1 \wedge a_4, a_2 \wedge a_5, a_3, a_6}(N)
     +S_{a_1 \wedge a_4, a_2 \wedge a_5, a_6, a_3}(N) \nonumber\\ &-&
      S_{a_1 \wedge a_4, a_2 \wedge a_5, a_3 \wedge a_6}(N)
     -S_{a_1, a_2, a_3}(N) S_{a_4, a_5, a_6}(N) =0~.
\end{eqnarray}
We consider algebraic relations of harmonic sums to express the harmonic sums of
a given depth in terms of a minimal set of harmonic sums. 
For larger depth
in addition to the
relation given above multiple shuffles may contribute in principle.
%%%%%%%%%%%%%%%%%%%%%%%%%%%%%%%%%%%%%%%%%%%%%%%%%%%%%%%%%%%%%%%%%%%%%%%%
\subsection{Integration-by-Parts Relations}
%%%%%%%%%%%%%%%%%%%%%%%%%%%%%%%%%%%%%%%%%%%%%%%%%%%%%%%%%%%%%%%%%%%%%%%%

\vspace{1mm}
\noindent
The finite harmonic sums are related to the harmonic 
polylogarithms~\cite{VR} by a weighted {\sc Mellin} transform up to terms 
of lower weight. 
The harmonic polylogarithms derive from the following three functions
%------------------------------------------------------------------------------$
\begin{eqnarray}
f(0;x)  &=& \frac{1}{x} \nonumber\\
f(1;x)  &=& \frac{1}{1-x} \nonumber\\
f(-1;x) &=& \frac{1}{1+x} 
\end{eqnarray}
%------------------------------------------------------------------------------
by iterated integrals~:
%------------------------------------------------------------------------------$
\begin{eqnarray}
H_a(x)  &=& \int_0^x dz f(a;z)  \nonumber\\
H_{\pm 1,b_1 \ldots b_m}(x) &=& \int_0^x dz f(\pm 1;z) H_{b_1, \ldots,
b_m}(z)~, \nonumber\\
H_{b_1 + 1, b_2, \ldots, b_m}(x) &=& \int_0^x dz f(0;z)H_{b_1, \ldots,
b_m}(z)~, 
\end{eqnarray}
%------------------------------------------------------------------------------
where in the set ${a,b_1, \ldots, b_m}$ not all indices are zero and 
%------------------------------------------------------------------------------$
\begin{eqnarray}
\left.
H_{a,b_1 \ldots b_m}(x)\right|_{a=0, b_i=0} = \frac{1}{(m+1)!} 
\ln^{m+1}(x)~.   
\end{eqnarray}
%------------------------------------------------------------------------------
Since harmonic polylogarithms $H_{a_1, \ldots, a_n}(x)$ obey the 
integration-by-parts relation
%------------------------------------------------------------------------------$
\begin{eqnarray}
\label{eqIBP}
H_{a_1, \ldots, a_n}(x) &=& H_{a_1}(x) H_{a_2, \ldots, a_n}(x)
-H_{a_2, a_1}(x) H_{a_3, \ldots, a_n}(x) \nonumber\\ & & 
+ \ldots + (-1)^{n+1} H_{a_n, \ldots, a_1}(x)
\end{eqnarray}
%------------------------------------------------------------------------------
they can potentially imply new relations between harmonic sums extending 
the number of relations which were discussed in the foregoing section.
Harmonic polylogarithms are somewhat simpler objects than harmonic sums.
Their algebraic product relation 
is
%------------------------------------------------------------------------------$
\begin{eqnarray}
\label{eqHPR}
H_{a_1, \ldots, a_n}(x) H_{b_1, \ldots, b_m}(x)&=&  
H_{a_1, \ldots, a_n}(x) \SH H_{b_1, \ldots, b_m}(x)~.
\end{eqnarray}
%------------------------------------------------------------------------------
Due to this it is evident that all relations derived in the forthcoming 
sections for harmonic sums turn into those for harmonic polylogarithms 
simply removing all terms containing the $\wedge$--symbol.

For $n=2$ (\ref{eqIBP}) and (\ref{eqHPR}) lead to the same relation. For 
$n=3$ one obtains out of a combination of the algebraic relations
(\ref{eqHPR}) that
%------------------------------------------------------------------------------$
\begin{eqnarray}
\label{eqREL3a}
H_{a_1}(x) H_{a_2,a_3}(x) - H_{a_3}(x) H_{a_2,a_1}(x) &=&  
H_{a_1}(x) \SH H_{a_2,a_3}(x) - H_{a_3}(x) \SH H_{a_2,a_1}(x) 
\end{eqnarray}  
%------------------------------------------------------------------------------
%------------------------------------------------------------------------------$
\begin{eqnarray}
\label{eqREL3b} 
H_{a_1,a_2,a_3}(x) = H_{a_1}(x) H_{a_2,a_3}(x) - H_{a_2,a_1}(x) H_{a_3}(x) 
+ H_{a_3,a_2, a_1}(x)   
\end{eqnarray}
%------------------------------------------------------------------------------
and therefore no new relations. The explicit consideration of the cases
$n=4$ to 6 yields no new relations as well. Binary shuffle products of
objects of length $n_1$ and $n_2$ contain $\binom{n_1+n_2}{n_1}$ summands.
Let us rewrite Eq.~(\ref{eqIBP}) subtracting the l.h.s. written as the 
first addend in the r.h.s. $H_{a_1, \ldots, a_n}$ is the shuffle product
of itself 
with ${\sf 1}$. The number of index--shuffled harmonic polylogarithms of 
depth $n$ with a positive sign are equal to those with a negative sign 
since 
%------------------------------------------------------------------------------$
\begin{eqnarray}
\sum_{k=0}^{n} \binom{n}{k}(-1)^{k+1} = 0~. 
\end{eqnarray}
%------------------------------------------------------------------------------
Combining always two consecutive summands in the rewritten form of 
(\ref{eqIBP}) from the left to the right, one finds that all shuffles
starting with $a_1$ are annihilated in the first pair. The remaining terms
are added to the term $H_{a_2, a_1}(x) H_{a_3, \ldots, a_{n}}(x)$, where
all shuffles starting with $a_2$ are annihilated etc. We conclude that 
the relations for harmonic sums which correspond to (\ref{eqIBP}) are 
fully contained in the relations derived in the foregoing section.

Finally we note two interesting properties of the (shuffle) product
of harmonic polylogarithms. We apply the differential and integral
operator, resp., onto the difference of the shuffle product and the
product of two harmonic polylogarithms.
%------------------------------------------------------------------------------  
\begin{eqnarray}
\label{eqHPRDI}
\frac{d}{dx} \left[-H_{a_1, \ldots, a_n}(x) H_{b_1, \ldots, b_m}(x)
+ H_{a_1, \ldots, a_n}(x) \SH H_{b_1, \ldots, b_m}(x)\right] &=& 
\nonumber\\
-f(a_1;x) \left[H_{a_2, \ldots, a_n}(x) H_{b_1, \ldots, b_m}(x)
- H_{a_2, \ldots, a_n}(x) \SH H_{b_1, \ldots, b_m}(x)\right] & &
\nonumber\\
-f(b_1;x) \left[ H_{a_1, \ldots, a_n}(x) H_{b_2, \ldots, b_m}(x)
- H_{a_1, \ldots, a_n}(x) \SH H_{b_2, \ldots, b_m}(x)\right]
&=&  0~.
\end{eqnarray}
%------------------------------------------------------------------------------
In (\ref{eqHPRDI}) it is assumed that the harmonic polylogarithm with no
indices corresponds to the unit element {\sf 1}.
Differentiation for $x$ maps the binary algebraic relations of degree
(depth) $n+m$ to those of degree $n+m-1$. Conversely the application of
the integral operator maps an algebraic relation of depth $n+m-1$ to a one
of degree $n+m$ integrating (\ref{eqHPRDI}) definitely,
%------------------------------------------------------------------------------$
\begin{eqnarray}
\label{eqHPRINT}
 \left[-H_{a_1, \ldots, a_n}(x) H_{b_1, \ldots, b_m}(x)
+ H_{a_1, \ldots, a_n}(x) \SH H_{b_1, \ldots, b_m}(x)\right] &=&
\nonumber\\
-\int_0^x dz f(a_1;z) \left[H_{a_2, \ldots, a_n}(z) H_{b_1, \ldots,
b_m}(z)
- H_{a_2, \ldots, a_n}(z) \SH H_{b_1, \ldots, b_m}(z)\right] & &
\nonumber\\ 
- \int_0^x dz f(b_1;z) \left[ H_{a_1, \ldots, a_n}(z) H_{b_2, \ldots,
b_m}(x)
- H_{a_1, \ldots, a_n}(z) \SH H_{b_2, \ldots, b_m}(z)\right]
&=&  0~.
\end{eqnarray}
%------------------------------------------------------------------------------
In this way differentiation and integration create downwards and upwards
moves in the tree of shuffles of harmonic polylogarithms which may be 
used in the sense outlined above as general representations for 
shuffles of any type. The connection between harmonic polylogarithms and
harmonic sums is easily established~\cite{VR} by the {\sc Mellin}
transform of the former 
%------------------------------------------------------------------------------$
\begin{eqnarray}
\label{eqMEL}
\MV[\left. H_{a_1, \ldots, a_n}(x) \right|_{reg}](N) =
\int_0^1 dx~x^{N-1}~\left. H_{a_1, \ldots, a_n}(x) \right|_{reg}   
\end{eqnarray}
%------------------------------------------------------------------------------
applying appropriate +-distribution regularizations, which are
obtained such that the integral (\ref{eqMEL}) exists. A unique definition
is achieved writing each finite harmonic sum in terms of a linear
combination of {\sc Mellin} transforms of harmonic polylogarithms. The
fact that the integration-by-parts relation does not result into new
relations for harmonic polylogarithms holds therefore also for the finite
harmonic sums.
%%%%%%%%%%%%%%%%%%%%%%%%%%%%%%%%%%%%%%%%%%%%%%%%%%%%%%%%%%%%%%%%%%%%%%%%
\subsection{Complete Permutations of the Index--Set}
%%%%%%%%%%%%%%%%%%%%%%%%%%%%%%%%%%%%%%%%%%%%%%%%%%%%%%%%%%%%%%%%%%%%%%%%
%

\vspace{1mm}
\noindent
We now consider the sums of all finite harmonic sums of a given rank 
over a complete permutation of the index--set. This combination of
harmonic 
sums can be represented as a polynomial of single harmonic sums and has 
therefore as well a simple analytic continuation to complex values of $N$ 
in terms of $\psi$--functions and their derivatives. Up to 
rank~4 the corresponding relations were given before~\cite{HS3} and read
%-----------------------------------------------------------------------
\begin{eqnarray}
S_{a_1,a_2} + S_{a_2,a_1} 
&=& S_{a_1} S_{a_2} + S_{a_1 \wedge a_2},~~~~~\mbox{\cite{EUL}}\\
\sum_{{\rm perm}\{a_1,a_2,a_3\}}
S_{a_1,a_2,a_3} 
&=& S_{a_1} S_{a_2} S_{a_3} +
\sum_{{\rm inv~perm}} S_{a_1} S_{a_2 \wedge a_3}
+ 2 S_{a_1 \wedge a_2 \wedge a_3},~~~~~\mbox{\cite{IND}}\\
\sum_{{\rm perm}\{a_1,a_2,a_3,a_4\}}
S_{a_1,a_2,a_3,a_4}
&=& S_{a_1} S_{a_2} S_{a_3} S_{a_4} +
\sum_{{\rm inv~perm}} S_{a_1} S_{a_2}
S_{a_3 \wedge a_4} \nonumber\\ & &
+
\sum_{{\rm inv~perm}} S_{a_1 \wedge a_2}
S_{a_3 \wedge a_4} + 2
\sum_{{\rm inv~perm}} S_{a_1}
S_{a_2 \wedge a_3 \wedge a_4} + 6
S_{a_1 \wedge a_2 \wedge a_3 \wedge a_4},~~~\mbox{\cite{HS3}}.\nonumber\\
\end{eqnarray}
%-----------------------------------------------------------------------
Here `perm' denotes all permutations and
{\rm `inv~perm'} denotes all permutations in which a single index
in a $\wedge$--contraction is only used once. 

A general way to derive these relations consists in summing over the 
general relation Eq.~(\ref{eqPROD1})
accounting for the weight factors. In the r.h.s. still polynomials
of harmonic sums of degree larger than one are contained which have
to be combined with the help of the complete permutation relations for
the sums of lower rank. Finally one obtains a polynomial out of
harmonic sums of rank one. The relations for the sums of depth~5 and 6
read~\footnote{These and the relations up to depth~10 were derived 
in~\cite{PWILL}.}
%-----------------------------------------------------------------------
\begin{eqnarray}
\sum_{{\rm perm}} S_{a_1,a_2,a_3,a_4,a_5}
&=& S_{a_1} S_{a_2} S_{a_3} S_{a_4} S_{a_5} 
+ \sum_{\rm inv~perm} S_{a_1 \wedge a_2} S_{a_3} S_{a_4} S_{a_5}
\nonumber\\  & &
+ \sum_{\rm inv~perm} S_{a_1 \wedge a_2} S_{a_3 \wedge a_4} S_{a_5}
+ 2 \sum_{\rm inv~perm} S_{a_1 \wedge a_2 \wedge a_3} S_{a_4} S_{a_5}
\nonumber\\  & &
+ 2 \sum_{\rm inv~perm} S_{a_1 \wedge a_2 \wedge a_3} S_{a_4 \wedge a_5}
+ 6 \sum_{\rm inv~perm} S_{a_1 \wedge a_2 \wedge a_3 \wedge a_4} S_{a_5}
\nonumber\\  & &
+ 24 \sum_{\rm inv~perm} S_{a_1 \wedge a_2 \wedge a_3 \wedge a_4 
\wedge a_5} \\
%----------------------------
\sum_{{\rm perm}} S_{a_1,a_2,a_3,a_4,a_5,a_6}
&=& S_{a_1} S_{a_2} S_{a_3} S_{a_4} S_{a_5} S_{a_6}
+ \sum_{\rm inv~perm} S_{a_1 \wedge a_2} S_{a_3} S_{a_4} S_{a_5} S_{a_6}
\nonumber\\  & &
+ \sum_{\rm inv~perm} S_{a_1 \wedge a_2} S_{a_3 \wedge a_4} 
S_{a_5}  S_{a_6}
+ \sum_{\rm inv~perm} S_{a_1 \wedge a_2} S_{a_3 \wedge a_4} 
S_{a_5 \wedge a_6} \nonumber \\ & &
+ 2 \sum_{\rm inv~perm} S_{a_1 \wedge a_2 \wedge a_3} 
S_{a_4} S_{a_5} S_{a_6}
+ 2 \sum_{\rm inv~perm} S_{a_1 \wedge a_2 \wedge a_3} S_{a_4 \wedge a_5}
S_{a_6} \nonumber\\ & &
+ 4 \sum_{\rm inv~perm} S_{a_1 \wedge a_2 \wedge a_3} 
S_{a_4 \wedge a_5 \wedge a_6}
+ 6 \sum_{\rm inv~perm} S_{a_1 \wedge a_2 \wedge a_3 \wedge a_4} 
S_{a_5} S_{a_6} \nonumber\\ & &
+ 6 \sum_{\rm inv~perm} S_{a_1 \wedge a_2 \wedge a_3 \wedge a_4} 
S_{a_5 \wedge a_6}
+ 24 \sum_{\rm inv~perm} S_{a_1 \wedge a_2 \wedge a_3 \wedge a_4 
\wedge a_5} S_{a_6} \nonumber \\ & &
+ 120 \sum_{\rm inv~perm} S_{a_1 \wedge a_2 \wedge a_3 \wedge a_4
\wedge a_5 \wedge a_6}~.
\end{eqnarray}
%-----------------------------------------------------------------------

The construction recipe for the general case is now straightforward.
Sum over all possible polynomial structures in $S_{p_i}, 
S_{p_i \wedge p_j}$~etc. We associate with an $l_1$-fold
$\wedge$-contraction the weight factor $l=(l_1 -1)!$.
The respective weight factors of the corresponding product of single sums 
is the product of the weights $l$. The sum over the complete permutation 
of a given index set is associated to symmetric 
polynomials~\cite{LW,SYPOL}.

From the above relations one obtains all $k$--fold finite harmonic sums
with a single index,
%-----------------------------------------------------------------------
\begin{eqnarray}
\label{eqS2}
S_{a,a}   &=& \frac{1}{2} \left[S_a^2 + S_{2a}\right]\\
S_{a,a,a} &=& \frac{1}{6} \left[S_a^3 + 3 S_a S_{2a} + 2 S_{3a} \right]\\
S_{a,a,a,a} &=& \frac{1}{24} \left[S_a^4 + 6 S_a^2 + S_{2a} + 3 S_{2a}^2
+ 8 S_a S_{3a} + 6 S_{4a} \right]\\
S_{a,a,a,a,a} &=& \frac{1}{120} \left[S_a^5 + 10 S_a^3 S_{2a} + 20 S_a^2
S_{3a} + 30 S_a S_{4a} + 15 S_a S_{2a}^2 + 20 S_{2a} S_{3a} + 24 S_{5a}
\right] \nonumber\\
\\
\label{eqS6}
S_{a,a,a,a,a,a} &=& \frac{1}{720} \left[S_a^6 + 15 S_{2a} S_a^4
+ 40 S_{3a} S_a^3 + 90 S_{4a} S_a^2 + 144 S_a S_{5a} + 45 S_a^2 S_{2a}^2
\right.
\nonumber\\ 
& & ~~~~\left.
+ 120 S_a S_{2a} S_{3a} + 15 S_{2a}^3 +90 S_{2a} S_{4a} + 40 S_{3a}^2
+ 120 S_{6a}\right]~.
\end{eqnarray}
%-----------------------------------------------------------------------
If $a < 0$, i.e. an alternating sum is considered, the symbol $na$ is
evaluated as $na = +|na|$ for $n$ even and $na = - |na|$ for $n$ odd.
Eqs.~(\ref{eqS2}--\ref{eqS6}) are equivalent to determinant relations
discussed in Ref.~\cite{HS3} and were studied in different contexts
as invariant theory  of algebraic equations of various variables and 
integer sums in Refs.~\cite{FDB,RAMAN} before. Similar type 
determinant--relations emerge in various aspects for symmetric 
polynomials, see e.g.~\cite{SYPOL}.

%%%%%%%%%%%%%%%%%%%%%%%%%%%%%%%%%%%%%%%%%%%%%%%%%%%%%%%%%%%%%%%%%%%%%%%%
\section{The threefold sums revisited}
%%%%%%%%%%%%%%%%%%%%%%%%%%%%%%%%%%%%%%%%%%%%%%%%%%%%%%%%%%%%%%%%%%%%%%%%
%

\vspace{1mm}
\noindent
In a previous investigation algebraic relations between the finite
non--alternating harmonic sums were obtained by decomposing the
sum
%-----------------------------------------------------------------------
\begin{equation}
T = \sum_{k=1}^N \frac{1}{k^a} \sum_{l=1}^{k} \frac{1}{l^b}
                               \sum_{m=1}^{k} \frac{1}{m^c}
\end{equation}
%-----------------------------------------------------------------------
by {\sc Borwein} and {\sc Girgensohn}~Ref.~\cite{GB}. $T$ can be
represented by
four different decompositions out of which three relations between the
finite harmonic sums $S_{a_1,a_2,a_3}$.~\footnote{For the Zeta--values
 the complete algebraic relations were quoted in \cite{GB}.} result.
These relations were extended
to the case of the finite harmonic sums with alternating indices in 
Ref.~\cite{HS3}
and read
%-----------------------------------------------------------------------
\begin{eqnarray}
\label{eq3A}
S_{a,b,c} &=& - S_{c,a,b} - S_{a,c,b} + S_c S_{a,b} + S_{c,a \wedge b}
- S_c S_{a \wedge b} + S_{a \wedge b, c} + S_{a \wedge c,b}
+S_{a, b \wedge c} - S_{a \wedge b \wedge c} \\
%-----------------------------------------------------------------------
S_{b,a,c} &=&   S_{c,a,b} - S_c S_{a,b} - S_{c,a \wedge b}
+ S_c S_{a \wedge b} + S_b S_{a,c} + S_{b,a \wedge c} - S_b S_{a \wedge
c}\\
%-----------------------------------------------------------------------
\label{eq3B}
S_{b,c,a} &=&  - S_{c,a,b} - S_{c,b,a}
+ S_c S_{a,b} + S_{c,a \wedge b}
- S_c S_{a \wedge b} + S_{b \wedge c,a} + S_c S_{b,a} - S_b S_{a,c}
+ S_b S_{a \wedge c}
\end{eqnarray}
%-----------------------------------------------------------------------
The partial permutation of indices induces the relation
%-----------------------------------------------------------------------
\begin{eqnarray}
S_{a,b,c} + S_{b,a,c} + S_{b,c,a} = S_a S_{b,c} + S_{a \wedge b,c}
+ S_{b, a \wedge  c}~.
\end{eqnarray}
%-----------------------------------------------------------------------
The index structure of this equation can now be permuted leading to six
equations. The coefficient matrix of this system of linear equations
%-----------------------------------------------------------------------
\begin{equation}
M_{3a} = \left\|
\begin{array}{cccccc} 
1 & 0 & 1 & 1 & 0 & 0 \\
0 & 1 & 0 & 0 & 1 & 1 \\
1 & 1 & 1 & 0 & 0 & 0 \\
0 & 0 & 0 & 1 & 1 & 1 \\
1 & 1 & 0 & 0 & 1 & 0 \\
0 & 0 & 1 & 1 & 0 & 1
\end{array} \right\|
\end{equation}
%-----------------------------------------------------------------------
is of rank~4. Due to this one more relation between the threefold
finite harmonic
sums than  found in~\cite{GB} is obtained in terms of polynomials of
harmonic sums of lower rank. All sums with three different indices
can be expressed by two chosen sums~:
%-----------------------------------------------------------------------
\begin{eqnarray}
\label{eq3C}
 S_{a,b,c} &=& S_c S_{a,b} + S_{a,b \wedge c}
-S_a S_{c,b}-S_{c,a \wedge b} +  S_{c,b,a}
\\
S_{a,c,b} &=&- S_b  S_{c,a}-     S_{b \wedge c,a}+
S_{b,c,a}+     S_{{a}}     S_{{c,b}}+     S_{{a \wedge  c,b}}
\\
S_{{b,a,c}} &=&     S_{{a \wedge b,c}}
-     S_{{c}}     S_{{a,b}}-     S_{{a,b \wedge c}}
+     S_{{a}}     S_{{b,c}}-     S_{{b,c,a}}+
     S_{{b,a \wedge c}}+       S_{{a}}     S_{{c,b}}+     
S_{{c,a \wedge b}}-
     S_{{c,b,a}} 
\\
\label{eq3D}
      S_{{c,a,b}} &=& S_{{b}}  S_{{c,a}}+  S_{{b \wedge c,a}}+
S_{{c,a \wedge b}}- S_{{b,c,a}}-  S_{{c,b,a}}~.
\end{eqnarray}
%-----------------------------------------------------------------------
The relations Eqs.~(\ref{eq3A}--\ref{eq3B}) are
contained in Eqs.~(\ref{eq3C}--\ref{eq3D}) for three different indices.
The corresponding harmonic sums up to weight~4 are expressed
by the sums
%-----------------------------------------------------------------------

\vspace{1mm}\noindent  % [-2,1,-1] [-1,-2,1]; [2,1,-1], [-1,2,1]
%{\sf Weight~4}
\begin{eqnarray}
S_{-1,1,-2} &=& S_{-2,1,-1} + S_{-2} S_{-1,1} + S_{-1,-3}
                -S_{-1} S_{-2,1} - S_{-2,-2} \\
S_{1,-2,-1} &=& S_{-1,-2,1} + S_1 S_{-2,-1} + S_{-3,-1} - S_{-1} S_{-2,1}
                -S_{3,1} \\
S_{-2,-1,1} &=& S_{3,1} + S_{1,3} - S_{-3,-1} - S_{-1,-3} +S_{-2,-2}
                -S_{1} S_{-1,-2} + 2 S_{-1} S_{-2,1}
                + S_{-1} S_{1,-2}\nonumber\\ & & 
- S_1 S_{-2,-1} - S_{-2,1,-1} -
S_{-1,-2,1}%\\  
\end{eqnarray}
\begin{eqnarray}
S_{1,-1,-2} &=& S_{-2} S_{1,-1} + S_{1,3} + S_{3,1}- S_{-2,1,-1} 
                - S_{-1,-2,1} - S_1 S_{-2,-1} + S_{-1} S_{-2,1}\\
S_{-1,1,2} &=&  S_2 S_{-1,1} + S_{-1,3} - S_{-1} S_{2,1} - S_{2,-2} + S_{2,1,-1}
\\
S_{1,2,-1} &=&  S_{-1,2,1} + S_1 S_{2,-1} + S_{3,-1} - S_{-1} S_{2,1} - S_{-3,1}
\\
S_{2,-1,1} &=&  S_{3,-1} + S_{2,-2}  - S_{2,1,-1}
               -S_{-1,2,1} - S_{3,-1} + S_{-1} S_{2,1} + S_{-3,1}
\\
S_{1,-1,2} &=&  -S_1 S_{2,-1} - S_{3,-1} + 2 S_{-1} S_{2,1} + S_{-3,1}
                -S_{-1,2,1}+S_{-2,2} - S_2 S_{-1,1} - S_{-1,3} \nonumber\\
                & & +S_{-1} S_{1,2} + S_{1,-3} + S_{2,-2} - S_{2,1,-1}~. 
\end{eqnarray}
%-----------------------------------------------------------------------
Comparing with the representations of the harmonic sums in terms of 
{\sc Mellin} transforms~Ref.~\cite{HS3},~Eq.~(39--102) one finds that
the {\sc Mellin} transform of the functions
%-----------------------------------------------------------------------
\begin{eqnarray}
\frac{\log(1-x)}{1+x} \Li_2(x) \hspace*{1cm} {\rm and}
\hspace*{1cm} \frac{\log(1+x)}{x-1} \Li_2(-x)
\end{eqnarray}
%-----------------------------------------------------------------------
earlier being
counted to the set of basic functions, cf.~\cite{HS3,ANCONT}, can
be expressed in terms of the algebraic relations~(\ref{eq3A}--\ref{eq3B}).
The set of basic functions needed up to two--loop order is thus reduced
to at most 23 functions.

At given depth $n$ the number of harmonic sums with  $k_i$ equal indices
such that $\sum_i k_i = n$ is
%-----------------------------------------------------------------------
\begin{equation}
\label{eqPERM}
n_{\rm perm}(\{a_1, \ldots, a_n\}) = \frac{n!}{\prod k_i!}~.
\end{equation}
%-----------------------------------------------------------------------
For threefold harmonic sums with two different indices three sums emerge
and the respective coefficient matrix reads
%-----------------------------------------------------------------------
\begin{equation}
M_{3b} = \left\|
\begin{array}{cccccc}
2 & 1 & 0 \\
0 & 1 & 2 \\
1 & 1 & 1  \\
\end{array} \right\|~.
\end{equation}
%----------------------------------------------------------------------- 
The corresponding system of linear equations 
is of rank~2. One obtains
%-----------------------------------------------------------------------
\begin{eqnarray}
S_{a,b,a} &=& 
- 2 S_{b,a,a} + S_{a} S_{b,a} + S_{a \wedge b,a} + S_{b,a \wedge a} \\
S_{a,a,b} &=& S_{b,a,a} 
           - \frac{1}{2} \left[S_a S_{b,a} + S_{a \wedge b,a}
           + S_{b, a\wedge a} - S_a S_{a,b} - S_{a \wedge a,b} - S_{a, a
\wedge b} \right]~,
\end{eqnarray}
%-----------------------------------------------------------------------
see Ref.~\cite{HS3}.

In summary the threefold harmonic sums are characterized as follows
\begin{center}
\renewcommand{\arraystretch}{1.3}
\begin{tabular}[h]{||l|c|c|c|c||}
\hline \hline %
\multicolumn{1}{||c|}{\sf Index Set}&
\multicolumn{1}{c|}{\sf Number}&
\multicolumn{1}{c|}{\sf Dep. Sums of Depth 3}&
\multicolumn{1}{c|}{\sf min. Weight}&
\multicolumn{1}{c||}{\sf Fraction of}\\
\multicolumn{1}{||c|}{\sf }&
\multicolumn{1}{c|}{\sf }&
\multicolumn{1}{c|}{\sf }&
\multicolumn{1}{c|}{\sf }&  
\multicolumn{1}{c||}{\sf fund. Sums}\\
\hline\hline  
$\{a,a,a\}$ &   1  &  1 & 3  &   0\\
\hline
$\{a,a,b\}$ &   3  &  2 & 3  &   1/3\\
\hline
$\{a,b,c\}$ &   6  &  4 & 4  &   1/3\\
\hline \hline
\end{tabular} 
\renewcommand{\arraystretch}{1.0}
\end{center}

%%%%%%%%%%%%%%%%%%%%%%%%%%%%%%%%%%%%%%%%%%%%%%%%%%%%%%%%%%%%%%%%%%%%%%%%
\section{The Fourfold Harmonic Sums}
%%%%%%%%%%%%%%%%%%%%%%%%%%%%%%%%%%%%%%%%%%%%%%%%%%%%%%%%%%%%%%%%%%%%%%%%
\renewcommand{\theequation}{\thesection.\arabic{equation}}
\setcounter{equation}{0}

\vspace{1mm}\nonumber
Five types of fourfold sums emerge. Their characteristics is summarized in 
the subsequent table.
\begin{center}
\renewcommand{\arraystretch}{1.3}
\begin{tabular}[h]{||l|c|c|c|c||}
\hline \hline
\multicolumn{1}{||c|}{\sf Index Set}&
\multicolumn{1}{c|}{\sf Number}&
\multicolumn{1}{c|}{\sf Dep. Sums of Depth 4}&
\multicolumn{1}{c|}{\sf min. Weight}&
\multicolumn{1}{c||}{\sf Fraction of}\\
\multicolumn{1}{||c|}{\sf }& %
\multicolumn{1}{c|}{\sf }&   %
\multicolumn{1}{c|}{\sf }&
\multicolumn{1}{c|}{\sf }&
\multicolumn{1}{c||}{\sf fund. Sums}\\
\hline\hline
$\{a,a,a,a\}$ &   1  &  1 & 4 & 0\\
\hline
$\{a,a,a,b\}$ &   4  &  3 & 4 & 1/4 \\
$\{a,a,b,b\}$ &   6  &  5 & 4 & 1/6 \\
\hline
$\{a,a,b,c\}$ &  12  &  9 & 5 & 1/4 \\
\hline
$\{a,b,c,d\}$ &  24  & 18 & 6 & 1/4\\
\hline \hline 
\end{tabular}
\renewcommand{\arraystretch}{1.0}
\end{center} 

%%%%%%%%%%%%%%%%%%%%%%%%%%%%%%%%%%%%%%%%%%%%%%%%%%%%%%%%%%%%%%%%%%%%%%%%
\subsection{Harmonic Sums with 4 Different Indices}
%%%%%%%%%%%%%%%%%%%%%%%%%%%%%%%%%%%%%%%%%%%%%%%%%%%%%%%%%%%%%%%%%%%%%%%%

\vspace{1mm}
\noindent
For this set of indices 24 different harmonic sums exist. As for the
threefold harmonic sums one may write down the associated system of
linear equations. The coefficient matrix has a size of $24 \times 48.$
It is obtained considering all index permutations for 
Eqs.~(\ref{eqR4A},\ref{eqR4B}). 
The rank of the coefficient
matrix is~18, i.e. 6 harmonic sums are chosen to express the remaining sums. 
Since none of the first 18 diagonal elements after bringing matrix  into
diagonal form vanishes we may use the last 6 sums as basic sums.  
Here and in the following we will not present
the respective coefficient matrices being too large in size.
One obtains the following relations~:
%--------------------------------------------------------------------------------
\begin{eqnarray}
%
% Arguments are:
%
%[d, a, b, c], [d, a, c, b], [d, b, a, c],
%[d, b, c, a], [d, c, a, b], [d, c, b, a]
%
S_{a,b,c,d} &=&
-S_{c} S_{a,d,b}-S_{c}S_{a,b,d}-S_{c \wedge d,a,b}-S_{c}S_{d,a,b}+S_{a}S_{d,c,b}
-S_{d,c,b,a}+S_{a \wedge d,b,c}\nonumber\\ & &
+S_{d}S_{a,b,c}+S_{a,b}S_{c,d}+S_{a \wedge d,c,b}-S_{a, c \wedge d,b}-S_{a,d,b \wedge c}
-S_{d,a,b \wedge c}-S_{a \wedge c,b \wedge d}\nonumber\\ & &
+S_{a,c,b \wedge d}+S_{c,a \wedge d,b}
+S_{c,a,b \wedge d}+S_{a,b \wedge d,c}+S_{d,c,a \wedge b}
%\end{eqnarray} \begin{eqnarray}
\\
%----------------------------------------------------------------------------------
S_{a,b,d,c} &=&
S_{c}S_{a,d,b}+S_{c}S_{a,b,d}+S_{c \wedge d,a,b}
+S_{c}S_{d,a,b}-S_{a}S_{d,c,b}+S_{d,c,b,a}-S_{a,b}S_{c,d} \nonumber\\ & &
+S_{d,a,b \wedge c}-S_{a \wedge d,b,c}
-S_{a \wedge d,c,b}+S_{a, c \wedge d,b}-S_{d,a \wedge b,c}
+S_{a,d,b \wedge c}+S_{a \wedge c,b \wedge d} \nonumber\\ & &
+S_{a,b,c \wedge d}-S_{a,c,b \wedge d}
-{S}_{c,a \wedge d,b}-S_{c,a,b \wedge d}-S_{d,b,a \wedge c}
-S_{a}S_{d,b,c}-S_{d,c,a \wedge b}
\nonumber\\ & &+S_{d,b,a,c}
+S_{d,b,c,a}
\\
%----------------------------------------------------------------------------------
S_{a,c,b,d} &=&
S_{d}S_{a,c,b}-S_{a,d,b \wedge c}+S_{a \wedge d,b \wedge c}-S_{d,a,b \wedge c}
-S_{a,c}S_{d,b}+S_{d,b,a \wedge c}+S_{a}S_{d,b,c} \nonumber\\ &&
+S_{a,c,b \wedge d}-S_{d,b,c,a}
\\
%----------------------------------------------------------------------------------
S_{a,c,d,b} &=&
-S_{a}S_{d,c,b}+S_{d,c,b,a}+S_{a,c \wedge d,b}
+S_{a,d,b \wedge c}-S_{a \wedge d,b \wedge c}+S_{d,a,b \wedge c}
+S_{a,c}S_{d,b} \nonumber\\ & &
-S_{d,a \wedge c,b}-S_{d,b,a \wedge c}
-S_{a}S_{d,b,c}-S_{d,c,a \wedge b}+S_{d,c,a,b}
+S_{d,b,c,a}
\\
%----------------------------------------------------------------------------------
S_{a,d,b,c} &=&
-S_{d,b,c,a}-S_{d,b,a,c}-S_{d,a,b,c}+S_{a}S_{d,b,c}
+S_{a \wedge d,b,c}+S_{d,a \wedge b,c}+S_{d,b,a \wedge c}
\\
%----------------------------------------------------------------------------------
S_{a,d,c,b} &=&
-S_{d,c,b,a}-S_{d,c,a,b}-S_{d,a,c,b}
+S_{a}S_{d,c,b}+S_{a \wedge d,c,b}+S_{d,a \wedge c,b}+S_{d,c,a \wedge b}
\\
%----------------------------------------------------------------------------------
S_{b,a,c,d} &=&
S_{c}S_{a,d,b}+S_{c}S_{a,b,d}+S_{c \wedge d,a,b}+S_{c}S_{d,a,b}-S_{d}S_{a,c,b}
-S_{b}S_{c,a,d}-S_{b}S_{c,d,a} \nonumber\\ & &
-S_{d}S_{a,b,c}-S_{a,b}S_{c,d}
+S_{d,a,b \wedge c}-S_{a \wedge d,b,c}-S_{a \wedge d,c,b}+S_{a}S_{b,c,d}+S_{a}S_{c,b,d}
\nonumber\\ & &
+S_{a,d,b \wedge c}+S_{d,a \wedge c,b}+S_{a \wedge c,b \wedge d}-S_{a,b \wedge d,c}
+{S}_{a}S_{c,d,b}+S_{c,b,a \wedge d}+S_{a \wedge c,d,b}
\nonumber\\ & &+S_{a \wedge b,c,d}+S_{b,a \wedge c,d}
+S_{b,c,a \wedge d}-2\,S_{a,c,b \wedge d}-2\,S_{c,a,b \wedge d}+S_{a \wedge c,b,d}
-S_{b \wedge c,a,d} \nonumber\\ & &
-S_{b \wedge c,d,a}-S_{c,b \wedge d,a}-S_{d,c,a,b} 
\\
%----------------------------------------------------------------------------------
S_{b,a,d,c} &=&
-S_{c}S_{a,d,b}-S_{c}S_{a,b,d}-S_{c \wedge d,a,b}-S_{c}S_{d,a,b}+S_{a,b}S_{c,d}
-S_{a,c \wedge d,b}+S_{b}S_{a,d,c} \nonumber\\ &&
-S_{d,a,b \wedge c}-S_{d,a \wedge c,b}
-S_{a \wedge c,b \wedge d}+S_{a,b \wedge d,c}-S_{a,b,c \wedge d}+S_{a \wedge b,d,c}
+S_{a,c,b \wedge d} \nonumber 
\end{eqnarray}\begin{eqnarray}
& &
+S_{c,a \wedge d,b}+S_{c,a,b \wedge d}+S_{d,a,b,c}
+S_{d,c,a,b}+S_{d,a,c,b}
\\
%----------------------------------------------------------------------------------
S_{b,c,a,d} &=&
S_{c}S_{a,d,b}+S_{c}S_{a,b,d}+S_{c \wedge d,a,b}+S_{c}S_{d,a,b}+S_{d}S_{a,c,b}
+S_{b}S_{c,a,d}+S_{b}S_{c,d,a} \nonumber\\ & &
-S_{a,b}S_{c,d}-S_{a \wedge d,b,c}
-S_{a}S_{b,c,d}-S_{a,b}S_{d,c}-{S}_{a}S_{c,b,d}+2\,S_{a,c \wedge d,b}
-S_{b}S_{a,d,c} 
\\  
& &
-S_{a,d,b \wedge c}+S_{a \wedge c,b \wedge d}-S_{a,b \wedge d,c}
+S_{a \wedge d,b \wedge c}
+S_{b,a,c \wedge d}-S_{a \wedge b,c \wedge d}-S_{a}S_{c,d,b}
\nonumber\\
& & +2\,S_{a,b,c \wedge d}
-S_{c,b,a \wedge d}
-S_{a \wedge c,d,b}-S_{b,c,a \wedge d}-2\,S_{c,a \wedge d,b}-S_{a \wedge c,b,d}
+S_{b \wedge c,a,d} \nonumber\\ & &
+S_{b \wedge c,d,a}+S_{c,b \wedge d,a}+S_{a,d}S_{b,c}-S_{d,a,c,b} 
\\
%----------------------------------------------------------------------------------
S_{d,b,a,c} &=&
-S_{c}S_{a,d,b}-S_{c}S_{a,b,d}-S_{c \wedge d,a,b}-S_{c}S_{d,a,b}-S_{a}S_{d,c,b}
+S_{d,c,b,a}+S_{a,b}S_{c,d} \nonumber\\ & &
+S_{a \wedge d,b,c}+S_{a}S_{b,c,d}
+S_{a,b}S_{d,c}-S_{a,c \wedge d,b}+S_{b}S_{a,d,c}+S_{a,d,b \wedge c}-S_{d,a \wedge c,b}
\nonumber\\ & &
-S_{a \wedge c,b \wedge d}+S_{a,b \wedge d,c}-S_{a \wedge d,b \wedge c}-S_{b,a,c \wedge d}
+S_{a \wedge b,c \wedge d}-2\,S_{a,b,c \wedge d}+S_{b,c,a \wedge d}
\nonumber\\ & &+S_{a,c,b \wedge d}
+S_{c,a \wedge d,b}+S_{c,a,b \wedge d}-S_{a,d}S_{b,c}-S_{d,c,a \wedge b}+S_{d,c,a,b}
+S_{d,a,c,b} \\
%----------------------------------------------------------------------------------
S_{b,d,a,c} &=&
S_{b \wedge d,a,c}+S_{b}S_{d,a,c}+S_{d,a,b \wedge c}+S_{d,a \wedge b,c}
-S_{d,a,b,c}-S_{d,a,c,b}-S_{d,b,a,c} 
\\
%----------------------------------------------------------------------------------
S_{b,d,c,a} &=&
-S_{d,c,b,a}-S_{d,c,a,b}-S_{d,b,c,a}-S_{b}S_{d,a,c}
-S_{b \wedge d,a,c}-S_{d,a,b \wedge c}-S_{b}S_{a,d,c}
\nonumber\\ & &-S_{a,b \wedge d,c}-S_{a,d,b \wedge c}
+S_{a}S_{b,d,c}+S_{b,a \wedge d,c}+S_{b,d,a \wedge c}+S_{a}S_{d,b,c}
+S_{a \wedge d,b,c}
\nonumber\\ & &+S_{d,b,a \wedge c}+S_{a}S_{d,c,b}+S_{a \wedge d,c,b}+S_{d,a \wedge 
c,b} +S_{d,c,a \wedge b} 
\\
%----------------------------------------------------------------------------------
S_{c,a,b,d} &=&
S_{c}S_{a,b,d}-S_{d}S_{a,c,b}+S_{a \wedge c,b,d}-S_{d}S_{a,b,c}+S_{a,d,b \wedge c}
-S_{a \wedge d,b \wedge c}+S_{d,a \wedge b,c}
\nonumber\\ & &+S_{d,a,b \wedge c}+S_{a,c}S_{d,b}
+S_{a,b \wedge c,d}-S_{a,b \wedge d,c}-S_{a,c,b \wedge d}-S_{d,b,a,c}
\\
%----------------------------------------------------------------------------------
S_{c,a,d,b} &=&
S_{c}S_{a,d,b}+S_{a \wedge c,d,b}-S_{a \wedge d,b,c}-S_{a \wedge d,c,b}
+S_{a \wedge d,b \wedge c}-S_{d,a \wedge b,c}-S_{d,a,b \wedge c}
\nonumber\\ & &-S_{a,c}S_{d,b}
+S_{d,a,b,c}+S_{d,a,c,b}+S_{d,b,a,c} 
\\
%----------------------------------------------------------------------------------
S_{c,b,a,d} &=&
-2\,S_{c}S_{a,d,b}-2\,S_{c}S_{a,b,d}-S_{c \wedge d,a,b}-S_{c}S_{d,a,b}
-S_{b}S_{c,d,a}+S_{d}S_{a,b,c}+S_{a,b}S_{c,d}
\nonumber\\ & &+2\,S_{a \wedge d,b,c}
+S_{a \wedge d,c,b}+S_{a}S_{b,c,d}+S_{a,b}S_{d,c}+{S}_{a}S_{c,b,d}
-2\,S_{a,c \wedge d,b}+S_{b}S_{a,d,c}
\nonumber\\ & &-S_{a \wedge c,b \wedge d}+2\,S_{a,b \wedge d,c}
-S_{a \wedge d,b \wedge c}-S_{b,a,c \wedge d}+S_{a \wedge b,c \wedge d}+S_{a}S_{c,d,b}
-2\,S_{a,b,c \wedge d}
\nonumber\\ & &+S_{c,a \wedge b,d}+S_{c,b,a \wedge d}+S_{b,c,a \wedge d}
-S_{a,b \wedge c,d}+S_{a,c,b \wedge d}+2\,S_{c,a \wedge d,b}+S_{c,a,b \wedge d}
\nonumber\\ & &
-S_{b \wedge c,d,a}-S_{c,b \wedge d,a}-S_{a,d}S_{b,c}-S_{d,a,b,c} 
%\end{eqnarray}  
%\begin{eqnarray}
\\
%----------------------------------------------------------------------------------
S_{c,b,d,a} &=& 
2\,S_{c}S_{a,d,b}+S_{c}S_{a,b,d}+S_{c \wedge d,a,b}+S_{c}S_{d,a,b}+S_{b}S_{c,d,a}
-S_{a,b} S_{c,d}-2\,S_{a \wedge d,b,c}
\nonumber\\ & &-S_{a \wedge d,c,b}-S_{a}S_{b,c,d}
-S_{a,b}S_{d,c}+2\,S_{a,c \wedge d,b}-S_{b} S_{a,d,c}-S_{d,a \wedge b,c}
+S_{a \wedge c,b \wedge d}
\nonumber\\ & &-S_{a,b \wedge d,c}+S_{a \wedge d,b \wedge c}+S_{b,a,c \wedge d}
-S_{a \wedge b,c \wedge d}-S_{a}S_{c,d,b}+2\,S_{a,b,c \wedge d}-S_{b,c,a \wedge d}
\nonumber\\ & &-S_{a,c,b \wedge d}-2\,S_{c,a \wedge d,b}-S_{c,a,b \wedge d}+S_{b
\wedge c,d,a}
+S_{c,b \wedge d,a}-S_{d,b,a \wedge c}-S_{a}S_{d,b,c}\nonumber\\ & &+S_{a,d}S_{b,c}
+S_{d,a,b,c}+S_{d,b,a,c}+S_{d,b,c,a} 
\\
%----------------------------------------------------------------------------------
S_{c,d,a,b} &=&
S_{c \wedge d,a,b}+S_{c}S_{d,a,b}+S_{d,a \wedge c,b}+S_{d,a,b \wedge c}
-S_{d,a,b,c}-S_{d,c,a,b}-S_{d,a,c,b} 
\\
%----------------------------------------------------------------------------------
S_{c,d,b,a} &=&
-S_{d,c,b,a}-S_{d,b,c,a}-S_{d,b,a,c}-S_{c}S_{d,a,b}
-S_{c \wedge d,a,b}-S_{d,a,b \wedge c}-S_{c}S_{a,d,b}\nonumber\\ & &-S_{a,c \wedge
d,b}
-S_{a,d,b \wedge c}
+S_{a}S_{c,d,b}+S_{a \wedge d,b,c}+S_{c,a \wedge d,b}+S_{c,d,a \wedge b}
+S_{a}S_{d,b,c}+S_{a \wedge d,c,b}
\nonumber\\ & &+S_{d,a \wedge b,c}+S_{d,b,a \wedge c}
+S_{a}S_{d,c,b}+S_{d,c,a \wedge b}~. 
%----------------------------------------------------------------------------------
\end{eqnarray}
%-----------------------------------------------------------------------
Harmonic sums of this type occur for the first time at
the level of weight~6.

%%%%%%%%%%%%%%%%%%%%%%%%%%%%%%%%%%%%%%%%%%%%%%%%%%%%%%%%%%%%%%%%%%%%%%%%
\subsection{Harmonic Sums with 3 Different Indices}
%%%%%%%%%%%%%%%%%%%%%%%%%%%%%%%%%%%%%%%%%%%%%%%%%%%%%%%%%%%%%%%%%%%%%%%%

%\vspace{1mm}
\noindent
This class contains 12 different sums. The coefficient matrix is $M_{4b}$ is
of rank 9 and again we may choose the last 3 harmonic sums to express the
remaining 9. The relations for the sums are
%-----------------------------------------------------------------------
\begin{eqnarray}
%
% Arguments are:
%
%[c, a, a, b],  [c, a, b, a], [c, b, a, a]
%
%
S_{a,a,b,c} &=&
\frac{1}{2} \left[S_{a,a \wedge c,b}-S_a S_{a,c,b}-S_a
S_{c,a,b}+S_a S_{c,b,a}-S_{c,a,a \wedge b} \right. 
\nonumber\\
& &
\left. +S_{a \wedge c,a,b}+S_{a \wedge c, b,a}
-S_{a \wedge a,c,b}
-S_{a,c,a \wedge b}+S_{c,b,a \wedge a}
-S_{c,a \wedge a,b}\right]+S_c S_{a,a,b} \nonumber
\end{eqnarray}
\begin{eqnarray}
& &
-S_{c,b,a,a}
+S_{a,a,b \wedge c}
+\frac{1}{2} S_{c, a \wedge b, a}
\\
%-----------------------------------------------------------------------
S_{a,a,c,b} &=&
-\frac{1}{2} S_{a \wedge c,a,b}+S_{c,b,a,a}+S_{c,a,b,a}+S_{c,a,a,b}
-\frac{1}{2}\left[S_{a}S_{c,b,a}+S_{a \wedge c,a,b}+S_{c,a \wedge b,a}
\right. \nonumber\\ & & \left. +
S_{c,b,a \wedge a}-S_{a}S_{a,c,b}-S_{a \wedge a,c,b}
-S_{a,a \wedge c,b}-S_{a,c,a \wedge b}+
S_{a}S_{c,a,b}+S_{c,a \wedge a,b}
\right. \nonumber\\ & &\left. +S_{c,a,a \wedge b}\right]
\\
%-----------------------------------------------------------------------
S_{a,b,a,c} &=&
S_{a \wedge a,b,c}+S_{a,b} S_{a,c}-S_{a,a \wedge c,b}-S_{a \wedge a,b \wedge
c}
-S_{a \wedge c,a,b}+S_{c,a,a \wedge b}+S_{c,a \wedge a,b} \nonumber\\ &&
+S_{a}S_{c,a,b}+S_{a,a \wedge b,c}+S_{a \wedge a,c,b}-2\,S_{c}S_{a,a,b}
-S_{c,a,b,a}
\\
%-----------------------------------------------------------------------
S_{a,b,c,a} &=&
S_{a}S_{a,b,c}+S_{a,b,a \wedge c}+S_{a}S_{a,c,b}
-S_{a,b} S_{a,c}+S_{a,c,a \wedge b}+S_{a \wedge a,b \wedge c}
-S_{a \wedge c,b,a}
\nonumber\\ & &-S_{c,a \wedge b,a}-2\,S_{a,a,b \wedge c}-S_a
 S_{c,b,a}-S_{c,b,a \wedge a}+S_{c,a,b,a}+2\,S_{c,b,a,a}
\\
%-----------------------------------------------------------------------
S_{a,c,a,b} &=&
-S_{c,a,b,a}-2\,S_{c,a,a,b}+S_{a}S_{c,a,b}+S_{a \wedge c,a,b}
+S_{c,a \wedge a,b}+S_{c,a,a \wedge b}\\
%-----------------------------------------------------------------------
S_{a,c,b,a} &=&
-2\,S_{c,b,a,a}-S_{c,a,b,a}+S_{a}S_{c,b,a}+S_{a \wedge c,
b,a}+S_{c,a \wedge b,a}+S_{c,b,a \wedge a}
\\
%-----------------------------------------------------------------------
S_{b,a,a,c} &=&-S_{c,a,a,b}-S_{a,b} S_{a,c}
+\frac{1}{2} \left[
S_{a} S_{a,b,c}+S_{a}S_{a,c,b}+S_{a}S_{b,a,c}-S_{a}S_{b,c,a}
-S_{a}S_{c,a,b}-S_{a}S_{c,b,a}\right]\nonumber\\ & &
+S_{b}S_{c,a,a}+S_{c}S_{a,a,b}-S_{a,a,b \wedge c}
-\frac{1}{2}\left[S_{c,a \wedge a,b}+S_{a \wedge a,c,b}+S_{c,b,a \wedge a}
-S_{a,c,a \wedge b}
\right. \nonumber\\ & & \left.
-S_{c,a \wedge b,a}-S_{c,a,a \wedge b}-S_{a,a \wedge
c,b}-S_{a \wedge c,a,b}+S_{a \wedge c,b,a}+S_{a,a \wedge b,c}+S_{a \wedge
a,b,c}\right] \nonumber \\ & &
+
S_{a \wedge a,b \wedge c}+\frac{1}{2}\left[S_{a,b,a \wedge c}-S_{b,c,a
\wedge a}
+S_{b,a \wedge a,c}+S_{a \wedge b,a,c}-S_{b,a \wedge c,a}
+S_{b,a,a \wedge c} \right. \nonumber\\ & & \left.
-S_{a \wedge b,c,a}\right]+S_{b \wedge c,a,a}
\\
%-----------------------------------------------------------------------
S_{b,a,c,a} &=&
-S_{a}S_{a,b,c}-S_{a,b,a \wedge c}-S_{a}S_{a,c,b}
+S_{a}S_{b,c,a}+2\,S_{c,a,a,b}+S_{a \wedge b,c,a}+S_{b,c,a \wedge
a}\nonumber\\
& &
+S_{a,b}S_{a,c}+S_{b,a \wedge c,a}-S_{a,c,a \wedge b}
-S_{a \wedge a,b \wedge c}+S_{a \wedge c,b,a}-S_{c,a \wedge b,a}   
-2\,S_{c,a,a \wedge b} \nonumber\\ & &
+2\,S_{a,a,b\wedge c}+S_{a}S_{c,b,a}
+S_{c,b,a \wedge a}-2\,S_{b \wedge c,a,a}-2\,S_{b}S_{c,a,a}+S_{c,a,b,a}
\\
%-----------------------------------------------------------------------
S_{b,c,a,a} &=&
-S_{c,b,a,a}-S_{c,a,b,a}-S_{c,a,a,b}+S_{b}S_{c,a,a}+S_{b \wedge 
c,a,a}+S_{c,a \wedge b, a}
+S_{c,a,a \wedge b}~.
%-----------------------------------------------------------------------
\end{eqnarray}
%-----------------------------------------------------------------------
\begin{eqnarray}
\label{eqM4b}
{\footnotesize
M_{4b} =
\left |\left|\begin {array}{cccccccccccc}
2&0&1&1&0&0&0&0&0&0&0&0\\
0&2&0&0&1&1&0&0&0&0&0&0\\  
0&0&1&0&0&0&2&1&0&0&0&0\\
0&0&0&1&0&0&0&1&2&0&0&0\\
0&0&0&0&1&0&0&0&0&2&1&0\\
0&0&0&0&0&1&0&0&0&0&1&2\\
1&1&1&0&0&0&1&0&0&0&0&0\\
0&0&0&1&1&1&0&1&0&0&0&0\\
0&0&0&0&0&0&0&0&1&1&1&1\\
1&1&0&0&1&0&0&0&0&1&0&0\\
0&0&1&1&0&1&0&0&0&0&1&0\\
0&0&0&0&0&0&1&1&1&0&0&1\\
1&0&1&1&0&0&1&1&1&0&0&0\\
0&1&0&0&1&1&0&0&0&1&1&1\\
2&2&1&0&1&0&0&0&0&0&0&0\\
0&0&0&1&1&1&0&0&0&2&1&0\\
2&2&1&0&1&0&0&0&0&0&0&0\\
0&0&1&1&0&1&2&1&0&0&0&0\\
0&0&1&1&0&1&2&1&0&0&0&0\\
0&0&0&0&0&0&0&1&2&0&1&2\\
1&0&1&1&0&0&1&1&1&0&0&0\\
0&0&0&1&1&1&0&0&0&2&1&0\\
0&0&0&0&0&0&0&1&2&0&1&2\\ 
0&1&0&0&1&1&0&0&0&1&1&1
\end {array}\right|\right|}
\end{eqnarray}
%----------------------------------------------------------------------
%%%%%%%%%%%%%%%%%%%%%%%%%%%%%%%%%%%%%%%%%%%%%%%%%%%%%%%%%%%%%%%%%%%%%%%%
\subsection{Harmonic Sums with 2 Different Indices}
%%%%%%%%%%%%%%%%%%%%%%%%%%%%%%%%%%%%%%%%%%%%%%%%%%%%%%%%%%%%%%%%%%%%%%%%
In this class all harmonic sums 
can be expressed by a single sum of the respective depth and weight.
%%%%%%%%%%%%%%%%%%%%%%%%%%%%%%%%%%%%%%%%%%%%%%%%%%%%%%%%%%%%%%%%%%%%%%%%
\subsubsection{Index-Set \boldmath{$\{a,a,b,b\}$}}
%%%%%%%%%%%%%%%%%%%%%%%%%%%%%%%%%%%%%%%%%%%%%%%%%%%%%%%%%%%%%%%%%%%%%%%%

\vspace{1mm}
\noindent
The class contains 6 different sums. The coefficient matrix results
from Eqs.~(\ref{eqR4A},\ref{eqR4B}) and is given by
%-----------------------------------------------------------------------
\begin{equation}
{\small M_{4c} = \left\|
\begin{array}{cccccc}
2 & 1 & 1 & 0 & 0 & 0 \\
0 & 0 & 1 & 2 & 1 & 0 \\
0 & 0 & 1 & 0 & 1 & 2 \\
2 & 1 & 0 & 1 & 0 & 0 \\
0 & 1 & 2 & 0 & 1 & 0 \\
0 & 0 & 0 & 1 & 1 & 2 \\
1 & 1 & 1 & 1 & 1 & 1 \\
4 & 2 & 0 & 0 & 0 & 0 \\
0 & 1 & 2 & 2 & 1 & 0 \\
0 & 1 & 2 & 2 & 1 & 0 \\
0 & 0 & 0 & 0 & 2 & 4 \\
1 & 1 & 1 & 1 & 1 & 1 \\
\end{array} \right\|}
\end{equation}
%-----------------------------------------------------------------------
It is of rank~5. The following relations are obtained~:
%-----------------------------------------------------------------------
\begin{eqnarray}
\label{eqAABB}
%
% expressed through: [b, b, a, a]
%
S_{a,a,b,b} &=& -S_{b,b,a,a} -\frac{1}{2} \left[
S_{a}S_{b,a,b}-S_{a}S_{b,b,a}-S_{b,b,a \wedge a}+S_{a,b,a \wedge b}
+S_{a \wedge a, b,b}+S_{a}S_{a,b,b} \right. \nonumber\\ & & \left.
+S_{b,a \wedge a,b}-S_{a \wedge b,a,b}-S_{a \wedge b,b,a}-S_{b,a \wedge b,a}+
S_{b,a,a \wedge b}-S_{a,a \wedge b,b}\right]+S_{a,a,b \wedge b}\nonumber\\ & &+
S_{b}S_{a,a,b}
\\
%-------------------------------------------------------------------------------------
S_{a,b,a,b} &=&
\frac{1}{2}S_{a,b}^2-2\,S_{b}S_{a,a,b}+2\,S_{a \wedge a,b,b}
-\frac{1}{2} S_{a \wedge a,b \wedge b}-S_{a \wedge b,a,b}-S_{a
,a,b \wedge b}+2\,S_{b,b,a,a}\nonumber\\ & &
+S_{a,b,a \wedge b}+S_{a}S_{a
,b,b}-S_{b,b,a \wedge a}-S_{a}S_{b,b,a}-S_{a \wedge b,b,a}
-S_{b,a \wedge b,a}+S_{a}S_{b,a,b} \nonumber\\ & &
+S_{b,a \wedge a,b}+
S_{b,a,a \wedge b}
\\
%-------------------------------------------------------------------------------------
S_{a,b,b,a} &=&
-S_{a,a,b \wedge b}-\frac{1}{2}S_{a,b}^{2}+S_{a,b,a \wedge b
}+\frac{1}{2} S_{a \wedge a,b \wedge b}+S_{a}S_{a,b,b}
\\
%-------------------------------------------------------------------------------------
S_{b,a,a,b} &=&
-\frac{1}{2} S_{a,b}^{2}+S_{b}S_{a,a,b}-S_{a \wedge a,
b,b}+\frac{1}{2} S_{a \wedge a,b \wedge b}+S_{a \wedge b,a,b}
\\
%-------------------------------------------------------------------------------------
\label{eqBABA}
S_{b,a,b,a} &=&
S_{a,a,b \wedge b}-2\, S_{b,b,a,a}+\frac{1}{2}\left[S_{a,b}^2-S_{a \wedge a, b
\wedge b}\right]-S_{a,b,a \wedge b}-S_{a}S_{a,
b,b}+S_{b,b,a \wedge a}\nonumber\\ & &
+S_{a}S_{b,b,a}+S_{a \wedge b,b,a}
+S_{b,a \wedge b,a}
%-------------------------------------------------------------------------------------
\end{eqnarray}
%-----------------------------------------------------------------------
Depending on the particular structure of the integrals emerging to a
given weight it may be useful to choose a different sum than used in
(\ref{eqAABB}--\ref{eqBABA}) to express the remaining sums.
Up to depth~4 only the weight~4
{\sc Mellin} transform of
%-----------------------------------------------------------------------
\begin{eqnarray}
\left[\frac{\log^3(1+x)}{x-1}\right]_+
\end{eqnarray}
%-----------------------------------------------------------------------
contributes. It is related to the harmonic sum $S_{-1,1,1,-1}(N)$.

%%%%%%%%%%%%%%%%%%%%%%%%%%%%%%%%%%%%%%%%%%%%%%%%%%%%%%%%%%%%%%%%%%%%%%%%
\subsubsection{Index-Set \boldmath{$\{a,a,a,b\}$}}
%%%%%%%%%%%%%%%%%%%%%%%%%%%%%%%%%%%%%%%%%%%%%%%%%%%%%%%%%%%%%%%%%%%%%%%%

\vspace{1mm}
\noindent
This class contains 4 different sums. The coefficient matrix reads
%-----------------------------------------------------------------------
\begin{equation}
M_{4b} = \left\|
\begin{array}{cccc} 
3 & 1 & 0 & 0 \\
0 & 2 & 2 & 0 \\     
0 & 0 & 1 & 1 \\     
1 & 1 & 1 & 1 \\
\end{array} \right\|
\end{equation}
%-----------------------------------------------------------------------
and is of rank~3. All sums can be represented in terms of one. 
The sums are given by
%-----------------------------------------------------------------------
\begin{eqnarray}
S_{a,b,a,a} &=& - 3 S_{b,a,a,a} + S_a S_{b,a,a} + S_{a \wedge b,a,a}
+S_{b,a \wedge a,a} + S_{b,a,a \wedge a}\\
S_{a,a,b,a} &=& \frac{1}{2} \left[S_a S_{a,b,a} + S_{a \wedge a,b,a} +
S_{a,a \wedge b,a} + S_{a,b,a \wedge a} \right] 
- S_{a,b,a,a} \\
S_{a,a,a,b} &=& \frac{1}{3} \left  [S_a S_{a,a,b} + S_{a \wedge a,a,b}
+ S_{a,a \wedge a,b} + S_{a,a,a \wedge b} -   S_{a,a,b,a} \right]
\end{eqnarray}
%-----------------------------------------------------------------------
recursively. Up to weight~4 eight sums emerge, which can be 
all
expressed in terms of {\sc Mellin} polynomials of functions of lower
depth and the {\sc Mellin} transforms of
%-----------------------------------------------------------------------
\begin{eqnarray}
\frac{\log^3(1+x)}{1+x} 
\end{eqnarray} \begin{eqnarray}
\label{e441}
\frac{1}{1+x} \left[\Li_2\left(\frac{1-x}{2}\right)\log\left(\frac{1+x}{4}
\right)+ 2 \Sf\left(\frac{1-x}{2}\right)\right]
- \frac{1}{1+x} \log(2)\log(1-x)\log\left(\frac{1+x}{2}\right)~.
\end{eqnarray}
%-----------------------------------------------------------------------
The former function is related to $S_{-1,1,1,1}(N)$ and the latter to
$S_{1,-1,-1,-1}(N)$.
Note also that the {\sc Mellin} transforms of the functions
$\log^3(1+x)/(1+x)$ and $\log^3(1-x)/(1+x)$ differ only by {\sc Mellin}
polynomials containing functions of lower weight and can 
be
related directly since the associated harmonic sums belong to the same
index class.

The functions which represent through {\sc Mellin} transforms
the fourfold harmonic sums of weight~4
depending only on one index (+1 or --1) up to functions of lower rank read~:
%-----------------------------------------------------------------------
\begin{eqnarray}
\left\{
\frac{1}{x-1} \left[\Li_2\left(\frac{1-x}{2}\right)\log\left(\frac{1+x}{4}
\right)+ 2 \Sf\left(\frac{1-x}{2}\right)\right]
- \frac{1}{x-1} \log(2)\log(1-x)\log\left(\frac{1+x}{2}\right)\right\}_+
\end{eqnarray}\begin{eqnarray}
\left[
\frac{\log^3(1-x)}{1-x}\right]_+~.
\end{eqnarray}
%-----------------------------------------------------------------------
It is interesting to note that the  {\sc Mellin} transforms of these functions 
can be expressed completely 
in terms of polynomials of
$\psi^k(N)$ and $\beta^{(k)}(N) = (1/2) d^k/dx^k
\left[\psi((1+x)/2)-\psi(x/2) \right]$ functions.
Finally we mention that 
instead of (\ref{e441}) one may use
%-----------------------------------------------------------------------  
\begin{eqnarray}
\frac{1}{1+x} 
\left[2\Li_3\left(\frac{1-x}{2}\right)-\ln(1-x)\Li_2\left(\frac{1-x}{2}\right)
\right]
\end{eqnarray}
%-----------------------------------------------------------------------
as basic function which corresponds to $S_{-1,-1,-1,1}(N)$ which yields 
a shorter expression.
%%%%%%%%%%%%%%%%%%%%%%%%%%%%%%%%%%%%%%%%%%%%%%%%%%%%%%%%%%%%%%%%%%%%%%%%
\section{The Fivefold Harmonic Sums}
%%%%%%%%%%%%%%%%%%%%%%%%%%%%%%%%%%%%%%%%%%%%%%%%%%%%%%%%%%%%%%%%%%%%%%%%
%

\vspace{1mm}
\noindent
In the Table below  we specify the different contributing index sets 
w.r.t. their multiplicity and the level of minimal depth at which they
occur.

\begin{center}
\renewcommand{\arraystretch}{1.3}
\begin{tabular}[h]{||l|c|c|c|c||}
\hline \hline
\multicolumn{1}{||c|}{\sf Index Set}&
\multicolumn{1}{c|}{\sf Number}&
\multicolumn{1}{c|}{\sf Dep. Sums of Depth 5}&
\multicolumn{1}{c|}{\sf min. Weight}&
\multicolumn{1}{c||}{\sf Fraction of}\\
\multicolumn{1}{||c|}{\sf }&
\multicolumn{1}{c|}{\sf }& 
\multicolumn{1}{c|}{\sf }&
\multicolumn{1}{c|}{\sf }&
\multicolumn{1}{c||}{\sf fund. Sums}\\
\hline\hline
$\{a,a,a,a,a\}$ &   1  & 1 & 5  & 0 \\
\hline
$\{a,a,a,a,b\}$ &   5  & 4 & 5  & 1/5\\
$\{a,a,a,b,b\}$ &  10  & 8 & 5  & 1/5\\
\hline
$\{a,a,a,b,c\}$ &  20  & 16 & 6 & 1/5\\
$\{a,a,b,b,c\}$ &  30  & 24 & 6 & 1/5\\
\hline
$\{a,a,b,c,d\}$ &  60  & 48 & 7 & 1/5\\
\hline
$\{a,b,c,d,e\}$ & 120  & 96 & 9 & 1/5 \\
\hline \hline
\end{tabular}
\renewcommand{\arraystretch}{1.0}
\end{center}

The harmonic sums of the type $S_{a,a,b,c,d}(N)$ and 
$S_{a,b,c,d,e}(N)$ do  contribute only at weight~7 or 9 and higher, 
respectively. Here we list only the number of of independent sums, which 
is 12 and 24. Since the expressions of the dependent sums in terms of the 
respective basis become voluminous beginning with the fivefold sums we 
will give the complete result only for the simpler cases and present
one representative relation for the complicated cases. The complete result
can be obtained instead from {\tt http://www.desy.de/\~{}blumlein}.
Likewise we will present the coefficient matrices only for the simpler
cases in explicit form.
%%%%%%%%%%%%%%%%%%%%%%%%%%%%%%%%%%%%%%%%%%%%%%%%%%%%%%%%%%%%%%%%%%%%%%%%
\subsection{Harmonic Sums with 3 Different Indices}
%%%%%%%%%%%%%%%%%%%%%%%%%%%%%%%%%%%%%%%%%%%%%%%%%%%%%%%%%%%%%%%%%%%%%%%%
%%%%%%%%%%%%%%%%%%%%%%%%%%%%%%%%%%%%%%%%%%%%%%%%%%%%%%%%%%%%%%%%%%%%%%%%
\subsubsection{Index-Set \boldmath{$\{a,a,b,b,c\}$}}
%%%%%%%%%%%%%%%%%%%%%%%%%%%%%%%%%%%%%%%%%%%%%%%%%%%%%%%%%%%%%%%%%%%%%%%%

\vspace{1mm}
\noindent
This class contains 30 sums. 
It turns out that six of the sums are independent. 
Due to the structure of 
the coefficient matrix one may choose the latter six sums.
As a typical example $S_{a,a,b,b,c}$ reads
%%%%%%%%%%%%%%%%%%%%%%%%%%%%%%%%%%%%%%%%%%%%%%%%%%%%%%%%%%%%%%%%%%%%%%%%
\begin{eqnarray}
%-----------------------------------------------------------------------
S_{a,a,b,b,c} &=&
-\frac{1}{2}\,{ S}_{a}{ S}_{a,b,c,b}-\frac{1}{2}\,{ S}_{a}{ S}_{b,c,a,b}
-\frac{1}{2}\,{ S}_{b,a,c,{ a \wedge b}}+\frac{1}{2}\,{ S}_{b,c,{ a \wedge b},a}
-\frac{1}{2}\,{ S}_{a,b,c,{ a \wedge b}}
\nonumber\\ & &
+\frac{1}{2}\,{ S}_{a,{ a \wedge b},c,b}-\frac{1}{2}\,{ S}_{b,{ a \wedge a},c,b}
+\frac{1}{2}\,{ S}_{b,c,b,{ a \wedge a}}-\frac{1}{2}\,{ S}_{c,a,b,{ a \wedge b}}
-\frac{1}{2}\,{ S}_{b,c,a,{ a \wedge b}}
\nonumber\\ & & 
-\frac{1}{2}\,{ S}_{{ a \wedge a},b,c,b}-\frac{1}{2}\,{ S}_{{ a \wedge a},c,b,b}
+\frac{1}{2}\,{ S}_{c,b,b,{ a \wedge a}}-\frac{1}{2}\,{ S}_{b,c,{ a \wedge a},b}
-\frac{1}{2}\,{ S}_{a}{ S}_{c,a,b,b}
\nonumber\\ & & 
+\frac{1}{2}\,{ S}_{c,b,{ a \wedge a},b}-\frac{1}{2}\,{ S}_{c,{ a \wedge a},b,b}
+{ S}_{b,a,a,{ b \wedge c}}-{ S}_{{ b \wedge c},b,a,a}
-{ S}_{{ a \wedge b},a,{ b \wedge c}}
\nonumber\\ & & 
-{ S}_{a,{ a \wedge b},{ b \wedge c}}+\frac{1}{2}\,{ S}_{{ a \wedge c},b,b,a}
-{ S}_{{ a \wedge b},{ a \wedge c},b}+\frac{1}{2}\,{ S}_{a,b,{ a \wedge c},b}
+\frac{1}{2}\,{ S}_{{ a \wedge c},b,a,b}
\nonumber\\ & & 
+\frac{1}{2}\,{ S}_{a,{ a \wedge c},b,b}+\frac{1}{2}\,{ S}_{b,{ a \wedge c},a,b}
+\frac{1}{2}\,{ S}_{b,{ a \wedge c},b,a}+\frac{1}{2}\,{ S}_{{ a \wedge b},c,b,a}
+\frac{1}{2}\,{ S}_{c,{ a \wedge b},a,b}
\nonumber\\
& & 
+\frac{1}{2}\,{ S}_{{ a \wedge b},a,c,b}+\frac{1}{2}\,{ S}_{{ a \wedge b},c,a,b}
+\frac{1}{2}\,{ S}_{c,{ a \wedge b},b,a}-\frac{1}{2}\,{ S}_{a,c,{ a \wedge b},b}
-\frac{1}{2}\,{ S}_{a,c,b,{ a \wedge b}}
\nonumber
\end{eqnarray}\begin{eqnarray}
& & 
-\frac{1}{2}\,{ S}_{c,b,{ a \wedge b},a}-\frac{1}{2}\,{ S}_{c,b,a,{ a \wedge b}}
+{ S}_{a,b,a,{ b \wedge c}}+{ S}_{a,a,{ b \wedge c},b}
-\frac{1}{2}\,{ S}_{a}{ S}_{a,c,b,b}
\nonumber\\ 
& & 
+\frac{1}{2}\,{ S}_{a}{ S}_{b,c,b,a}-{ S}_{c,{ b \wedge b},a,a}
+{ S}_{a,a,b,{ b \wedge c}}+{ S}_{b,c}{ S}_{a,a,b}
-\frac{1}{2}\,{ S}_{a}{ S}_{b,a,c,b}
%\nonumber 
%\end{eqnarray} \begin{eqnarray}
\\
& & 
-\frac{1}{2}\,{ S}_{c,a,{ a \wedge b},b}+\frac{1}{2}\,{ S}_{b,a,{ a \wedge c},b}
+\frac{1}{2}\,{ S}_{{ a \wedge c},a,b,b}+\frac{1}{2}\,{ S}_{a}{ S}_{c,b,b,a}
+\frac{1}{2}\,{ S}_{a}{ S}_{c,b,a,b}
\nonumber\\ & & 
-{ S}_{b}{ S}_{a,a,b,c}-{ S}_{b}{ S}_{c,b,a,a}+{ S}_{c}{ S}_{a,a,b,b}
+{S}_{c,b,b,a,a}
\end{eqnarray}
%------------------------------------------------------------------------
%%%%%%%%%%%%%%%%%%%%%%%%%%%%%%%%%%%%%%%%%%%%%%%%%%%%%%%%%%%%%%%%%%%%%%%%%%%%%%%$
%%%%%%%%%%%%%%%%%%%%%%%%%%%%%%%%%%%%%%%%%%%%%%%%%%%%%%%%%%%%%%%%%%%%%%%%
\subsubsection{Index-Set \boldmath{$\{a,a,a,b,c\}$}}
%%%%%%%%%%%%%%%%%%%%%%%%%%%%%%%%%%%%%%%%%%%%%%%%%%%%%%%%%%%%%%%%%%%%%%%%

\vspace{1mm}
\noindent
This class contains 20 sums. 
Four sums are independent. The expression for $S_{a,a,a,b,c}$ is shown as
an example.
%-------------------------------------------------------------------------
%%%%%%%%%%%%%%%%%%%%%%%%%%%%%%%%%%%%%%%%%%%%%%%%%%%%%%%%%%%%%%%%%%%%%%%%
\begin{eqnarray}
S_{a,a,a,b,c} &=&
{  S}_{c}{  S}_{a,a,a,b}+{  S}_{c,b,a,a,a}+\frac{1}{6}\,{  S}_{a}{  S}_{a,c,b,a}
+\frac{1}{6}\,{  S}_{a}{  S}_{c,a,b,a}-\frac{1}{3}\,{  S}_{a}{  S}_{c,b,a,a}
\nonumber\\ & &
-\frac{1}{3} \left[{  S}_{a}{  S}_{c,a,a,b}+{  S}_{a}{  S}_{a,c,a,b}
+{  S}_{a}{  S}_{a,a,c,b}+{  S}_{c,b,{  a \wedge a},a}\right]
+\frac{1}{6}\,{  S}_{{  a \wedge a},c,b,a}
\nonumber\\ & &
-\frac{1}{3} \left[{  S}_{{  a \wedge a},c,a,b}
+{  S}_{c,b,a,{  a \wedge a}}+{  S}_{a,{  a \wedge a},c,b}
+{  S}_{{  a \wedge a},a,c,b}+{  S}_{c,{  a \wedge a},a,b}\right]
\nonumber\\ & &
+\frac{1}{6}\,{  S}_{c,{  a \wedge a},b,a}-\frac{1}{3}\left[{  S}_{c,a,{  a \wedge a},b}
+{  S}_{a,c,{  a \wedge a},b}\right]+\frac{1}{6}\left[{  S}_{a,c,b,{  a \wedge a}}
+{  S}_{c,a,b,{  a \wedge a}}\right]
\nonumber\\ & &
-\frac{1}{3}\,{  S}_{c,{  a \wedge b},a,a}
+\frac{1}{6}\,{  S}_{a,c,{ a \wedge b},a}
-\frac{1}{3}\left[{  S}_{a,c,a,{a \wedge b}}
+{  S}_{c,a,a,{  a \wedge b}}+{  S}_{a,a,c,{a \wedge b}}\right]
\nonumber\\ 
& &
+\frac{1}{6}\left[{  S}_{c,a,{  a \wedge b},a}+{  S}_{a,{  a \wedge c},b,a}\right]
+\frac{1}{3}\left[2{  S}_{a,{  a \wedge c},a,b}-{  S}_{{  a \wedge c},b,a,a}
+2{  S}_{{  a \wedge c},a,a,b}\right]
\nonumber\\
& &
+\frac{1}{6}\,{  S}_{{  a \wedge c},a,b,a}
+\frac{2}{3}\,{  S}_{a,a,{  a \wedge c},b}
+{  S}_{a,a,a,{  b \wedge c}}
\end{eqnarray}
%%%%%%%%%%%%%%%%%%%%%%%%%%%%%%%%%%%%%%%%%%%%%%%%%%%%%%%%%%%%%%%%%%%%%%%%%%%%%%%$
%%%%%%%%%%%%%%%%%%%%%%%%%%%%%%%%%%%%%%%%%%%%%%%%%%%%%%%%%%%%%%%%%%%%%%%%
\subsection{Harmonic Sums with 2 Different Indices}
%%%%%%%%%%%%%%%%%%%%%%%%%%%%%%%%%%%%%%%%%%%%%%%%%%%%%%%%%%%%%%%%%%%%%%%%
%%%%%%%%%%%%%%%%%%%%%%%%%%%%%%%%%%%%%%%%%%%%%%%%%%%%%%%%%%%%%%%%%%%%%%%%
\subsubsection{Index-Set \boldmath{$\{a,a,a,b,b\}$}}
%%%%%%%%%%%%%%%%%%%%%%%%%%%%%%%%%%%%%%%%%%%%%%%%%%%%%%%%%%%%%%%%%%%%%%%%

\vspace{1mm}
\noindent
This class contains 10 sums. The coefficient matrix given 
in Eq.~(\ref{eqM5c}) 
is of rank~8. All sums can be represented in terms of two. For
$S_{a,a,a,b,b}$ one obtains~:
%-----------------------------------------------------------------------
\begin{eqnarray}
S_{a,a,a,b,b}
&=&
{S}_{b,b,a,a,a}+{S}_{b}{S}_{a,a,a,b}+\frac{1}{6}\,{S}_{a}{S}_{a,b,b,a}
-\frac{1}{3}\,{S}_{a}{S}_{b,b,a,a}-\frac{1}{3}\,{S}_{a}{S}_{a,a,b,b}
\nonumber\\ & &
-\frac{1}{3}\,{S}_{a}{S}_{a,b,a,b}-\frac{1}{3}\,{S}_{a}{S}_{b,a,a,b}+\frac{1}{6}\,{S}_{a}{S}_{b,a,b,a}
-\frac{1}{3}\,{S}_{a \wedge a,a,b,b}-\frac{1}{3}\,{S}_{a,a \wedge a,b,b}
\nonumber\\ & &
+\frac{1}{6}\,{S}_{a \wedge a,b,b,a}-\frac{1}{3}\,{S}_{b,a,a \wedge a,b}+\frac{1}{6}\,{S}_{a,b,b,a \wedge a}
-\frac{1}{3}\,{S}_{a \wedge a,b,a,b}-\frac{1}{3}\,{S}_{b,a \wedge a,a,b}
\nonumber\\ & &
+\frac{1}{6}\,{S}_{b,a \wedge a,b,a}-\frac{1}{3}\,{S}_{a,b,a \wedge a,b}-\frac{1}{3}\,{S}_{b,b,a,a \wedge a}
+\frac{1}{6}\,{S}_{b,a,b,a \wedge a}-\frac{1}{3}\,{S}_{b,b,a \wedge a,a}
\nonumber\\ & &
+\frac{2}{3}\,{S}_{a,a,a \wedge b,b}-\frac{1}{3}\,{S}_{a,a,b,a \wedge
b}-\frac{1}{3}\,{S}_{a,b,a,a \wedge b}
+\frac{2}{3}\,{S}_{a,a \wedge b,a,b}+\frac{1}{6}\,{S}_{a,a \wedge b,b,a}
\nonumber\\ & &
-\frac{1}{3}\,{S}_{b,a,a,a \wedge b}+\frac{1}{6}\,{S}_{a,b,a \wedge b,a}-\frac{1}{3}\,{S}_{a \wedge b,b,a,a}
-\frac{1}{3}\,{S}_{b,a \wedge b,a,a}+\frac{2}{3}\,{S}_{a \wedge b,a,a,b}
\nonumber\\ & &
+\frac{1}{6}\,{S}_{a \wedge b,a,b,a}+\frac{1}{6}\,{S}_{b,a,a \wedge b,a}+{S}_{a,a,a,b \wedge b}
\end{eqnarray}
%-----------------------------------------------------------------------
%-----------------------------------------------------------------------
\begin{equation}
\label{eqM5c}
M_{5c} = \left\|
\begin{array}{cccccccccc}
                   3&1&1&0&0&0&0&0&0&0\\
                   0&2&0&2&1&0&0&0&0&0\\
                   0&0&2&0&1&2&0&0&0&0\\
                   0&0&0&1&0&0&3&1&0&0\\
                   0&0&0&0&1&0&0&2&2&0\\
                   0&0&0&0&0&1&0&0&1&3\\
                   2&1&0&1&0&0&1&0&0&0\\
                   0&1&2&0&1&0&0&1&0&0\\
                   0&0&0&1&1&2&0&0&1&0\\
                   0&0&0&0&0&0&1&1&1&2\\
                   3&2&2&1&1&1&0&0&0&0\\
                   0&1&0&1&1&0&4&2&1&0\\
                   0&0&1&0&1&1&0&1&3&3\\
                   6&3&0&1&0&0&0&0&0&0\\
                   0&2&4&2&2&0&0&0&0&0\\
                   0&0&0&1&0&2&3&3&1&0\\
                   0&1&2&3&1&0&2&1&0&0\\
                   0&0&0&0&2&5&0&2&1&0\\
                   0&0&0&0&0&0&0&1&3&6\\
                   1&1&1&1&2&1&1&0&1&1\\
\end{array} \right\|
\end{equation}
%-----------------------------------------------------------------------
%%%%%%%%%%%%%%%%%%%%%%%%%%%%%%%%%%%%%%%%%%%%%%%%%%%%%%%%%%%%%%%%%%%%%%%%
\subsubsection{Index-Set \boldmath{$\{a,a,a,a,b\}$}}
%%%%%%%%%%%%%%%%%%%%%%%%%%%%%%%%%%%%%%%%%%%%%%%%%%%%%%%%%%%%%%%%%%%%%%%%

\vspace{1mm}
\noindent
Finally this class contains 5 different sums. The coefficient matrix is
%-----------------------------------------------------------------------
\begin{equation}
M_{5d} = \left\|
\begin{array}{ccccc}
4 & 1 & 0 & 0 & 0\\
0 & 3 & 2 & 0 & 0\\
0 & 0 & 2 & 3 & 0\\
0 & 0 & 0 & 1 & 4\\
1 & 1 & 1 & 1 & 1\\
\end{array} \right\|
\end{equation}
%-----------------------------------------------------------------------
and has rank~4.
The sums are given by
%-----------------------------------------------------------------------
\begin{eqnarray}
S_{a,a,a,a,b} &=&
S_{b,a,a,a,a}-\frac{1}{4} \left[S_{a}S_{b,a,a,a}
+S_{a \wedge b,a,a,a}+S_{b,a \wedge a,a,a}+S_{b,a,a \wedge a,a}
+S_{b,a,a,a \wedge a}\right] \nonumber\\ & &
+\frac{1}{12}\left[S_{a} 
S_{a,b,a,a} 
+S_{a \wedge a,b,a,a}+S_{a, a \wedge b ,a,a}
+S_{a,b,a \wedge a,a}+S_{a,b,a,a \wedge a}
-S_{a}S_{a,a,b,a} \right. \nonumber\\ & & \left.
-S_{a \wedge a,a,b,a}-S_{a,a \wedge a,b,a} 
-S_{a,a,a \wedge b,a}
-S_{a,a,b,a \wedge a}\right]+\frac{1}{4}\left[S_{a}S_{a,a,a,b}+
S_{a \wedge a,a,a,b} \right. \nonumber\\ & & \left.
+S_{a,a \wedge a,a,b}
+S_{a,a,a \wedge a,b}+S_{a,a,a,a \wedge b}\right]
\\
%-------------------------------------------------------------------------
S_{a,a,a,b,a} &=&
-4 S_{b,a,a,a,a}+S_{a}S_{b,a,a,a}+S_{a \wedge b,a,a,a}
+S_{b,a \wedge a,a,a}+S_{b,a,a \wedge a,a}
+S_{b,a,a,a \wedge a}- \nonumber\\ & &
\frac{1}{3} \left[S_{a}S_{a,b,a,a} 
+S_{a \wedge a,b,a,a}
+S_{a,a \wedge b,a,a}+S_{a,b,a \wedge a,a}
+S_{a,b,a,a \wedge a}-S_a S_{a,a,b,a} \right. \nonumber\\ & & \left.
-S_{a \wedge a,a,b,a}
-S_{a,a \wedge a,b,a} 
-S_{a,a,a \wedge b,a}-S_{a,a,b,a \wedge a} \right]
\\   
%-------------------------------------------------------------------------
S_{a,a,b,a,a} &=&
6 S_{b,a,a,a,a}-\frac{3}{2} \left[S_{a}S_{b,a,a,a}+S_{a \wedge 
b,a,a,a}+S_{b,a \wedge a,a,a}+S_{b,a,a \wedge a,a}
+S_{b,a,a,a \wedge a}\right] \nonumber\\ & &
+\frac{1}{2} \left[S_{a}
S_{a,b,a,a}+S_{a \wedge a,b,a,a}+S_{a, a \wedge b,a,a}+
S_{a,b,a \wedge a,a}+S_{a,b,a,a \wedge a} \right]
\\   
%-------------------------------------------------------------------------
S_{a,b,a,a,a} &=&
-4 S_{b,a,a,a,a}+S_{a}S_{b,a,a,a}+S_{a \wedge b,a,a,a}+S_{b,a \wedge a,a,a}
+S_{b,a,a \wedge a,a}+S_{b,a,a,a \wedge a}
%-------------------------------------------------------------------------
\end{eqnarray}
%-----------------------------------------------------------------------

%%%%%%%%%%%%%%%%%%%%%%%%%%%%%%%%%%%%%%%%%%%%%%%%%%%%%%%%%%%%%%%%%%%%%%%%
\section{The Sixfold Harmonic Sums}
%%%%%%%%%%%%%%%%%%%%%%%%%%%%%%%%%%%%%%%%%%%%%%%%%%%%%%%%%%%%%%%%%%%%%%%

\vspace{1mm}
\noindent
The structure of the sixfold harmonic sums is summarized in the following 
Table. There are six basic classes of sums with up to three cases each.
{\small
\begin{center}
\renewcommand{\arraystretch}{1.3}
\begin{tabular}[h]{||l|c|c|c|c||}
\hline \hline
\multicolumn{1}{||c|}{\sf Index Set}&
\multicolumn{1}{c|}{\sf Number}&
\multicolumn{1}{c|}{\sf Dep. Sums of Depth 6}&
\multicolumn{1}{c|}{\sf min. Weight}&
\multicolumn{1}{c||}{\sf Fraction of}\\
\multicolumn{1}{||c|}{\sf }&
\multicolumn{1}{c|}{\sf }&
\multicolumn{1}{c|}{\sf }&
\multicolumn{1}{c|}{\sf }&
\multicolumn{1}{c||}{\sf fund. Sums}\\
\hline\hline
$\{a,a,a,a,a,a\}$ &   1 &   1 &  6 & 0   \\
\hline
$\{a,a,a,a,a,b\}$ &   6 &   5 &  6 & 1/6 \\
$\{a,a,a,a,b,b\}$ &  15 &  13 &  6 & 2/15\\
$\{a,a,a,b,b,b\}$ &  20 &  17 &  6 & 3/20\\
\hline
$\{a,a,a,a,b,c\}$ &  30 &  25 &  7 & 1/6 \\
$\{a,a,a,b,b,c\}$ &  60 &  50 &  7 & 1/6 \\
$\{a,a,b,b,c,c\}$ &  90 &  76 &  8 & 7/45\\
\hline
$\{a,a,a,b,c,d\}$ & 120 & 100 &  8 & {1}/{6} \\
$\{a,a,b,b,c,d\}$ & 180 & 150 &  8 & {1}/{6} \\
\hline
$\{a,a,b,c,d,e\}$ & 360 & 300 & 10 & {1}/{6} \\
\hline 
$\{a,b,c,d,e,f\}$ & 720 & 600 & 12 & {1}/{6} \\
\hline \hline
\end{tabular}
\renewcommand{\arraystretch}{1.0}
\end{center}
}

%%%%%%%%%%%%%%%%%%%%%%%%%%%%%%%%%%%%%%%%%%%%%%%%%%%%%%%%%%%%%%%%%%%%%%%%
\subsection{Harmonic Sums with 2 Different Indices} 
%%%%%%%%%%%%%%%%%%%%%%%%%%%%%%%%%%%%%%%%%%%%%%%%%%%%%%%%%%%%%%%%%%%%%%%%

%%%%%%%%%%%%%%%%%%%%%%%%%%%%%%%%%%%%%%%%%%%%%%%%%%%%%%%%%%%%%%%%%%%%%%%%
\subsubsection{Index-Set \boldmath{$\{a,a,a,b,b,b\}$}}
%%%%%%%%%%%%%%%%%%%%%%%%%%%%%%%%%%%%%%%%%%%%%%%%%%%%%%%%%%%%%%%%%%%%%%%%

\vspace{1mm} \noindent
This class contains 20 different sums which can be expressed by three 
basic sums. 
%The coefficient matrix is obtained as stack matrix out of 
%the matrices $M_{6a1}$ to $M_{6a3}$ in Appendix~A. 
As an example one obtains for $S_{a,a,a,b,b,b}$ 
%%%%%%%%%%%%%%%%%%%%%%%%%%%%%%%%%%%%%%%%%%%%%%%%%%%%%%%%%%%%%%%%%
%=====================================================================================
\begin{eqnarray}
%%%%
%  1
%%%%
S_{a,a,a,b,b,b} &=&
       \frac{1}{3} S_{a} S_{b, a, b, a, b} - \frac{1}{6} S_{b, a \wedge a, b, b, a}
     - \frac{1}{6} S_{a, b, b, b, a \wedge a} + \frac{1}{3} S_{b, a \wedge a, b, a, b}
     - \frac{1}{6} S_{a \wedge a, b, b, b, a} 
\nonumber\\ & &
     + \frac{1}{3} S_{a \wedge a, b, a, b, b}
     + \frac{1}{3} S_{a \wedge a, b, b, a, b} + \frac{1}{3} S_{a, b, b, a \wedge a, b}
     + \frac{1}{3} S_{a \wedge a, a, b, b, b} + \frac{1}{3} S_{a, b, a \wedge a, b, b}
\nonumber\\ & &
     - \frac{1}{6} S_{a, b, a, a \wedge b, b} - \frac{1}{6} S_{b, a \wedge b, a, b, a}
     - \frac{1}{6} S_{a \wedge b, b, a, a, b} - \frac{1}{6} S_{b, a \wedge b, a, a, b}
     + \frac{1}{3} S_{a \wedge b, b, b, a, a} 
\nonumber\\ & &
     + \frac{1}{3} S_{a, a \wedge b, a, b, b}
     - \frac{1}{6} S_{a, a, b, a \wedge b, b} + \frac{1}{3} S_{a, a, b, b, a \wedge b}
     + \frac{1}{3} S_{b, a, a \wedge a, b, b} + \frac{1}{3} S_{a} S_{a, a, b, b, b}
\nonumber\\ & &
     + \frac{1}{3} S_{b, b, a \wedge b, a, a} + \frac{1}{3} S_{b, a \wedge b, b, a, a}
     - \frac{1}{6} S_{a, b, b, a \wedge b, a} + \frac{1}{3} S_{b, a, b, a, a \wedge b}
     - \frac{1}{6} S_{a, a \wedge b, b, b, a} 
\nonumber\\ & &
     - \frac{1}{6} S_{a, b, a \wedge b, b, a}
     - \frac{1}{6} S_{b, a, a \wedge b, a, b} - \frac{1}{6} S_{a, a \wedge b, b, a, b}
     - \frac{1}{6} S_{a, b, a \wedge b, a, b} - \frac{1}{6} S_{a \wedge b, a, b, a, b}
\nonumber\\ & &
     + \frac{1}{3} S_{a, b, b, a, a \wedge b} - \frac{1}{2} S_{a, b \wedge b, a, a, b}
     - \frac{1}{2} S_{b \wedge b, a, a, a, b} - \frac{1}{2} S_{a, b, a, a, b \wedge b}
     - \frac{1}{2} S_{a, a, b \wedge b, a, b} 
\nonumber\\ & &
     - \frac{1}{6} S_{b, a, a \wedge b, b, a}
     + \frac{1}{3} S_{b, a, a, b, a \wedge b} + \frac{1}{3} S_{a, a, a \wedge b, b, b}
     + \frac{1}{3} S_{a, b, a, b, a \wedge b} - \frac{1}{6} S_{b, b, a, a \wedge b, a}
\nonumber\\ & &
     - \frac{1}{6} S_{a \wedge b, b, a, b, a} + \frac{1}{3} S_{b, b, a, a, a \wedge b}
     - \frac{1}{6} S_{b, a, a, a \wedge b, b} - \frac{1}{6} S_{b, a, b, a \wedge b, a}
     - \frac{1}{6} S_{b, b, a, b, a \wedge a} 
\nonumber\\ & &
     - \frac{1}{2} S_{b, a, a, a, b \wedge b}
     - \frac{1}{6} S_{a \wedge b, a, b, b, a} + \frac{1}{3} S_{b, b, a, a \wedge a, b}
     + \frac{1}{3} S_{b, b, b, a \wedge a, a} + \frac{1}{3} S_{b, b, b, a, a \wedge a}
\nonumber
\end{eqnarray}\begin{eqnarray} & &
     - \frac{1}{6} S_{b, b, a \wedge a, b, a} + \frac{1}{3} S_{b, a, b, a \wedge a, b}
     + \frac{1}{3} S_{a, a \wedge a, b, b, b} + \frac{1}{3} S_{b, b, a \wedge a, a, b}
     + \frac{1}{3} S_{b, a \wedge a, a, b, b} 
\nonumber\\ & &
     - \frac{1}{2} S_{a, a, b, a, b \wedge b}
     - \frac{1}{6} S_{b, a, b, b, a \wedge a} - S_{b, b, b, a, a, a} 
     + \frac{1}{3} S_{a \wedge b, a, a, b, b} + \frac{1}{3} S_{a} S_{b, b, a, a, b}
\nonumber\\ & &
     - \frac{1}{6} S_{a} S_{a, b, b, b, a} - \frac{1}{6} S_{a} S_{b, b, a, b, a}
     + \frac{1}{3} S_{a} S_{b, a, a, b, b} + \frac{1}{3} S_{a} S_{a, b, a, b, b}
     - \frac{1}{6} S_{a} S_{b, a, b, b, a} 
\nonumber\\ & &
     - \frac{1}{2} S_{b} S_{a, a, b, a, b}
     - \frac{1}{2} S_{b} S_{a, b, a, a, b} + \frac{1}{3} S_{a} S_{b, b, b, a, a}
     - \frac{1}{2} S_{b} S_{b, a, a, a, b} + \frac{1}{3} S_{a} S_{a, b, b, a, b}
\end{eqnarray}
%=====================================================================================
%%%%%%%%%%%%%%%%%%%%%%%%%%%%%%%%%%%%%%%%%%%%%%%%%%%%%%%%%%%%%%%%%%%%%%

%%%%%%%%%%%%%%%%%%%%%%%%%%%%%%%%%%%%%%%%%%%%%%%%%%%%%%%%%%%%%%%%%%%%%%%%
\subsubsection{Index-Set \boldmath{$\{a,a,a,a,b,b\}$}}
%%%%%%%%%%%%%%%%%%%%%%%%%%%%%%%%%%%%%%%%%%%%%%%%%%%%%%%%%%%%%%%%%%%%%%%%

\vspace{1mm} \noindent
This class contains 15 different sums which are represented by two basic 
sums. For $S_{a,a,a,a,b,b}$ one obtains
%%%%%%%%%%%%%%%%%%%%%%%%%%%%%%%%%%%%%%%%%%%%%%%%%%%%%%%%%%%%%%%%%%%%%eqna
\begin{eqnarray}
%%%%%
%%  1
%%%%%
S_{a,a,a,a,b,b} 
&=& -\frac{1}{4} S_{a} S_{b, a, a, a, b} + \frac{3}{4} S_{a \wedge b, a, a, a, b}
     - \frac{1}{4} S_{b, a, a, a, a \wedge b} + \frac{1}{12} S_{a} S_{a, a, b, b, a}
     + S_{a, a, a, a, b \wedge b} 
     \nonumber\\ & &
     - \frac{1}{12} S_{a \wedge a, b, b, a, a}
     - \frac{1}{12} S_{a, b, b, a \wedge a, a} - \frac{1}{12} S_{a, b, b, a, a \wedge a}
     - \frac{1}{4} S_{b, a \wedge a, a, a, b} - \frac{1}{4} S_{b, a, a \wedge a, a, b}
     \nonumber\\ & &
     - \frac{1}{4} S_{b, a, a, a \wedge a, b} - \frac{1}{4} S_{a, a \wedge a, a, b, b}
     - \frac{1}{4} S_{a, a, a \wedge a, b, b} - \frac{1}{4} S_{a, b, a \wedge a, a, b}
     - \frac{1}{4} S_{a, b, a, a \wedge a, b} 
     \nonumber\\ & &
     + \frac{1}{12} S_{a \wedge a, b, a, b, a}
     + \frac{1}{12} S_{a, b, a \wedge a, b, a} - \frac{1}{4} S_{a, a, a, b, a \wedge b}
     - \frac{1}{4} S_{a, a, b, a, a \wedge b} + \frac{1}{12} S_{a, a, a \wedge b, b, a}
     \nonumber\\ & &
     + \frac{3}{4} S_{a, a, a \wedge b, a, b} 
     - S_{b, b, a, a, a, a} + \frac{1}{4} S_{b, b, a, a \wedge a, a}
     - \frac{1}{4} S_{a \wedge a, a, a, b, b} + \frac{1}{12} S_{a, b, a, b, a \wedge a}
     \nonumber\\ & &
     + \frac{1}{12} S_{b, a \wedge a, a, b, a} 
     + \frac{1}{12} S_{b, a, a \wedge a, b, a}
     + \frac{1}{12} S_{b, a, a, b, a \wedge a} - \frac{1}{12} S_{b, a \wedge a, b, a, a}
     - \frac{1}{12} S_{b, a, b, a \wedge a, a} 
     \nonumber\\ & &
     + \frac{1}{4} S_{b, b, a \wedge a, a, a}
     + \frac{1}{4} S_{b, b, a, a, a \wedge a} - \frac{1}{4} S_{a \wedge a, a, b, a, b}
     - \frac{1}{4} S_{a, a \wedge a, b, a, b} - \frac{1}{4} S_{a, a, b, a \wedge a, b}
     \nonumber\\ & &
     + \frac{1}{12} S_{a \wedge a, a, b, b, a} 
     + \frac{1}{12} S_{a, a \wedge a, b, b, a}
     + \frac{1}{12} S_{a, a, b, b, a \wedge a} - \frac{1}{4} S_{a \wedge a, b, a, a, b}
     - \frac{1}{12} S_{b, a, b, a, a \wedge a} 
      \nonumber\\ & &
     + \frac{1}{12} S_{a \wedge b, a, a, b, a}
     + \frac{1}{12} S_{b, a, a, a \wedge b, a} - \frac{1}{12} S_{a \wedge b, a, b, a, a}
     + \frac{1}{4} S_{a \wedge b, b, a, a, a} + \frac{1}{4} S_{b, a \wedge b, a, a, a}
     \nonumber\\ & &
     - \frac{1}{12} S_{b, a, a \wedge b, a, a} 
      + \frac{3}{4} S_{a, a, a, a \wedge b, b}
     + \frac{1}{12} S_{a, a, b, a \wedge b, a} + \frac{3}{4} S_{a, a \wedge b, a, a, b}
     - \frac{1}{4} S_{a, b, a, a, a \wedge b} 
\nonumber\\
& &
     \nonumber\\ & &
     + \frac{1}{12} S_{a, a \wedge b, a, b, a}
     + \frac{1}{12} S_{a, b, a, a \wedge b, a} - \frac{1}{12} S_{a, a \wedge b, b, a, a}
     - \frac{1}{12} S_{a, b, a \wedge b, a, a} - \frac{1}{4} S_{a} S_{a, a, a, b, b}
     \nonumber\\ & &
     - \frac{1}{12} S_{a} S_{a, b, b, a, a} 
     + \frac{1}{12} S_{a} S_{a, b, a, b, a}
     - \frac{1}{12} S_{a} S_{b, a, b, a, a} + \frac{1}{12} S_{a} S_{b, a, a, b, a}
     - \frac{1}{4} S_{a} S_{a, b, a, a, b} 
     \nonumber\\ & &
     + S_{b} S_{a, a, a, a, b}
     + \frac{1}{4} S_{a} S_{b, b, a, a, a} - \frac{1}{4} S_{a} S_{a, a, b, a, b}
\end{eqnarray}
%------------------------------------------------------------------------------------------
%%%%%%%%%%%%%%%%%%%%%%%%%%%%%%%%%%%%%%%%%%%%%%%%%%%%%%%%%%%%%%%%%%%%%%%%
\subsubsection{Index-Set \boldmath{$\{a,a,a,a,a,b\}$}}
%%%%%%%%%%%%%%%%%%%%%%%%%%%%%%%%%%%%%%%%%%%%%%%%%%%%%%%%%%%%%%%%%%%%%%%%

\vspace{1mm}
\noindent
This class contains 6 different sums. The coefficient matrix reads
%-----------------------------------------------------------------------
\begin{equation}
M_{6c} = \left\|
\begin{array}{cccccc}
5 & 1 & 0 & 0 & 0 & 0\\
0 & 4 & 2 & 0 & 0 & 0\\
0 & 0 & 3 & 3 & 0 & 0\\
0 & 0 & 0 & 2 & 4 & 0\\
0 & 0 & 0 & 0 & 1 & 5\\
1 & 1 & 1 & 1 & 1 & 1\\
\end{array} \right\|
\end{equation}
%-----------------------------------------------------------------------
and is of rank~5.  All sums can be represented in terms of one.
The sums are given by
%-----------------------------------------------------------------------
\begin{eqnarray}
S_{a,a,a,a,a,b} &=&
-{ S}_{b,a,a,a,a,a}+\frac{1}{5}\,{ S}_{a}{ S}_{a,a,a,a,b}
+\frac{1}{30}\,{S}_{a}{ S}_{a,a,b,a,a}-\frac{1}{20}\left[{ S}_{a}{ S}_{a,a,a,b,a}
\right. \nonumber\\  & & \left.
+{ S}_{{ a \wedge a},b,a,a,a}\right]+\frac{1}{5}\left[{ S}_{b,{ a \wedge a},a,a,a}
+{ S}_{b,a,{ a \wedge a},a,a}+{ S}_{b,a,a,a,{ a \wedge a}} \right.
\nonumber\\ & & \left.
+{ S}_{a,a,a,a,{ a \wedge b}}\right]
+\frac{1}{30}\,{ S}_{a,a,{ a \wedge b},a,a}
-\frac{1}{20}\,{ S}_{a,a,a,{ a \wedge b},a}+\frac{1}{30}\,{ S}_{{ a \wedge a},a,b,a,a}
\nonumber\\ & &
+\frac{1}{5}\left[{ S}_{a,a,{ a \wedge a},a,b}+{ S}_{a,a,a,{ a \wedge a},b}
+{S}_{{ a \wedge a},a,a,a,b}
+{ S}_{a,{ a \wedge a},a,a,b} \right]
\nonumber\\ & &
+\frac{1}{30}\left[{ S}_{a,a,b,{ a \wedge a},a}
+{ S}_{a,a,b,a,{ a \wedge a}}\right]+\frac{1}{5}\left[{ S}_{a}{ S}_{b,a,a,a,a}
+{ S}_{b,a,a,{ a \wedge a},a}\right]
\nonumber\\ & &
-\frac{1}{20}\left[{ S}_{a,a,{ a \wedge a},b,a}+{ S}_{a,a,a,b,{ a \wedge a}}
+{ S}_{{ a \wedge a},a,a,b,a}+{ S}_{a,{ a \wedge a},a,b,a}\right.
\nonumber\\ & & \left.
+{ S}_{a,b,{ a \wedge a},a,a}
+{ S}_{a,b,a,{ a \wedge a},a}+{ S}_{a,b,a,a,{ a \wedge a}}
+{ S}_{a}{ S}_{a,b,a,a,a}\right.
\nonumber\\ & & \left.
+{ S}_{a,{ a \wedge b},a,a,a} \right]
+\frac{1}{5}\,{ S}_{{ a \wedge b},a,a,a,a}
+\frac{1}{30}\,{ S}_{a,{ a \wedge a},b,a,a}
\\
%-----------------------------------------------------------------------
S_{a,a,a,a,b,a} &=&
5\,{ S}_{b,a,a,a,a,a}-\frac{1}{6}\,{ S}_{a}{ S}_{a,a,b,a,a}
+\frac{1}{4}\left[{ S}_{a}{ S}_{a,a,a,b,a}+{ S}_{{ a \wedge a},b,a,a,a}\right]
-{ S}_{b,{ a \wedge a},a,a,a}
\nonumber\\ & &
-{ S}_{b,a,{ a \wedge a},a,a}-{ S}_{b,a,a,a,{ a \wedge a}}
-\frac{1}{6}{ S}_{a,a,{ a \wedge b},a,a}+\frac{1}{4}\,{ S}_{a,a,a,{ a \wedge b},a}
-\frac{1}{6}\left[{ S}_{{ a \wedge a},a,b,a,a} \right.
\nonumber\\ & & \left.
+{ S}_{a,a,b,{a \wedge a},a}+{ S}_{a,a,b,a,{ a \wedge a}}\right]
-{ S}_{a}{ S}_{b,a,a,a,a}-{ S}_{b,a,a,{ a \wedge a},a}
+\frac{1}{4}\left[{ S}_{a,a,{ a \wedge a},b,a} \right.
\nonumber\\ & & \left.
+{ S}_{a,a,a,b,{ a \wedge a}}+{ S}_{{ a \wedge a},a,a,b,a}
+{ S}_{a,{ a \wedge a},a,b,a}+{ S}_{a,b,{ a \wedge a},a,a} \right.
\nonumber\\ & & \left.
+{S}_{a,b,a,{ a \wedge a},a}
+{ S}_{a,b,a,a,{ a \wedge a}}+{ S}_{a}{ S}_{a,b,a,a,a}
+{ S}_{a,{ a \wedge b},a,a,a} \right]
\nonumber\\ & &
-{ S}_{{ a \wedge b},a,a,a,a}
-\frac{1}{6}\,{ S}_{a,{ a \wedge a},b,a,a}
\\
%-----------------------------------------------------------------------
S_{a,a,a,b,a,a} &=&
-10\,{ S}_{b,a,a,a,a,a}+2\,{ S}_{a}{ S}_{b,a,a,a,a}
+2\left[{ S}_{{ a \wedge b},a,a,a,a}+{ S}_{b,{ a \wedge a},a,a,a}
+{ S}_{b,a,{ a \wedge a},a,a} \right.
\nonumber\\ & & \left.
+{ S}_{b,a,a,{ a \wedge a},a}+{ S}_{b,a,a,a,{ a \wedge a}} \right]
-\frac{1}{2}\left[{ S}_{a}{ S}_{a,b,a,a,a}+{ S}_{{a \wedge a},b,a,a,a}
+{ S}_{a,{ a \wedge b},a,a,a} \right.
\nonumber\\ & & \left.
+{ S}_{a,b,{a \wedge a},a,a}+{ S}_{a,b,a,{ a \wedge a},a}
+{ S}_{a,b,a,a,{a \wedge a}}\right]+\frac{1}{3}\left[{ S}_{a}{ S}_{a,a,b,a,a}
\right. \nonumber\\ & & \left.
+{ S}_{{ a \wedge a},a,b,a,a}
+{ S}_{a,{ a \wedge a},b,a,a}+{ S}_{a,a,{ a \wedge b},a,a}
+{ S}_{a,a,b,{ a \wedge a},a} 
+{ S}_{a,a,b,a,{ a \wedge a}} \right]
\\
%-----------------------------------------------------------------------
S_{a,a,b,a,a,a} &=&
10\,{ S}_{b,a,a,a,a,a}-2\left[{ S}_{a}{ S}_{b,a,a,a,a}
+{ S}_{{ a \wedge b},a,a,a,a}+{ S}_{b,{ a \wedge a},a,a,a}
+{ S}_{b,a,{ a \wedge a},a,a}\right.
\nonumber\\ & & \left.
+{ S}_{b,a,a,{ a \wedge a},a}+{ S}_{b,a,a,a,{ a \wedge a}}\right]
+\frac{1}{2}\left[{ S}_{a}{ S}_{a,b,a,a,a}+{ S}_{{a \wedge a},b,a,a,a}
+{ S}_{a,{ a \wedge b},a,a,a} \right.
\nonumber\\ & & \left.
+{ S}_{a,b,{a \wedge a},a,a}+{ S}_{a,b,a,{ a \wedge a},a}
+{ S}_{a,b,a,a,{a \wedge a}} \right]
\end{eqnarray}\begin{eqnarray}
%-----------------------------------------------------------------------
S_{a,b,a,a,a,a} &=&
-5\,{ S}_{b,a,a,a,a,a}+{ S}_{a}{ S}_{b,a,a,a,a}
+{ S}_{{ a \wedge b},a,a,a,a}+{ S}_{b,{ a \wedge a},a,a,a}
+{ S}_{b,a,{ a \wedge a},a,a}
\nonumber\\ & &
+{ S}_{b,a,a,{ a \wedge a},a}+{ S}_{b,a,a,a,{ a \wedge a}}
\end{eqnarray}

Finally we consider the case of general depth $d$ for the harmonic sums of the
index set
${a,a, \ldots, a,b}$ and their respective permutations. They depend on
$1/d \cdot n_{\rm perm}$ fundamental sums, i.e. {\sf one sum}, only.
The coefficient matrix reads
%-----------------------------------------------------------------------
\begin{equation}
\label{eqMD}
M_{d;1} = \left\|
\begin{array}{ccccccc}
d-1 & 1 & 0 & 0 & 0 & \ldots &0\\
0 & d-2 & 2 & 0 & 0 & \ldots &0\\
0 & 0 & d-3 & 3 & 0 & \ldots & 0\\  
\vdots & \vdots & \vdots & \vdots & \vdots & \ldots &\vdots\\
0 & 0 & 0 & \ldots &0 & 1 & d-1\\
1 & 1 & 1 & \ldots  & 1 & 1 & 1\\
\end{array} \right\|~.
\end{equation}
%-----------------------------------------------------------------------
This matrix is of rank $d-1$, which may be easily seen adding the first
line to the $-(d-1)$-fold of the last line, etc. etc., and therefore
$d-1$ harmonic sums of this type are dependent, i.e. the fraction of
independent sums is $1/d$. In the foregoing tables we calculated the
fraction of independent sums out of all possible sums of a given index  
class up to depth $d=6$. For these cases
%-----------------------------------------------------------------------
\begin{equation}
\label{eqFRAC}
\frac{n_{\rm independent}}{n_{\rm perm}} \leq \frac{1}{d}~.
\end{equation}
%-----------------------------------------------------------------------
%%%%%%%%%%%%%%%%%%%%%%%%%%%%%%%%%%%%%%%%%%%%%%%%%%%%%%%%%%%%%%%%%%%%%%%%
\section{Number of Independent Harmonic Sums and the Number of Lyndon 
Words}
%%%%%%%%%%%%%%%%%%%%%%%%%%%%%%%%%%%%%%%%%%%%%%%%%%%%%%%%%%%%%%%%%%%%%%%

\vspace{1mm}
\noindent
The algebraic relations between the harmonic sums associated to a given 
index set studied in the present paper are induced by the multiplication 
relation between the sums independently of their specific value or 
structure otherwise. Therefore the foregoing results hold for all 
mathematical objects which obey these conditions. Since the multiplication 
relation (\ref{eqPROD}) is directly associated to the shuffle product 
$\SH$ one may determine the number of basic sums by means of mathematical 
relations derived for shuffle algebras. It turns out that the number of 
basic harmonic
sums of a given index set is given by the number of {\sc
Lyndon}~\footnote{I thank S. Weinzierl for hinting me to {\sc Lyndon} 
words.} words
which can be formed by the letters being contained in this set.
The sets of characters (words) being 
considered in the following are always understood as representatives for 
which we consider all permutations, as before, to find the maximal set of 
algebraic relations of harmonic sums. We will make use of results obtained
in the theory of words~\cite{LOTH} and free algebras~\cite{KAS,REUT}.

Let ${\mathfrak A} = \{a,b,c,d, \ldots\}$ be a finite alphabet. $a, b, c, d,
\ldots$ are called the letters of this alphabet and any sequence
$\{a_1, \ldots, a_k\}$ with $a_i~\epsilon~{\mathfrak A}$ is a word. For 
brevity one often writes this sequence as a non--commutative
product $a_1 a_2 \ldots a_k$, the {\sf concatenation product}.
The length of a word is the number of its 
letters, which corresponds to the {\sf depth} $d$ of the respective 
harmonic sums. 
${\mathfrak A}^*$ denotes the {\sf free monoid} on ${\mathfrak A}$ made
out of all words $w$ including the empty word {\sf 1}. The set of
non--empty words is denoted as ${\mathfrak A}^+$.
The alphabet ${\mathfrak A}$ is ordered by
%------------------------------------------------------------------------
\begin{equation}
\label{eqORD}
a < b < c < d < \ldots~.
\end{equation}
%------------------------------------------------------------------------
For a word 
%------------------------------------------------------------------------
\begin{eqnarray}
\label{eqWO}
w = p x s
\end{eqnarray}
%------------------------------------------------------------------------
any non--empty factor $p$ is called a {\sf prefix} 
and any non-empty factor $s$ a {\sf suffix} of $w$. In extension of 
(\ref{eqORD}) the order relation $<$ applies also to words. 
%------------------------------------------------------------------------
\begin{eqnarray}
\label{eqORD1}   
{\rm i)}& &~~u < v \Leftrightarrow  v = u x,~~~~x~\epsilon~{\mathfrak A}^+
~{\rm or~}\\
{\rm ii)}& &~~u = x a u', v = x b v', a < b, c, u', v' 
~\epsilon~{\mathfrak A}^+~.
\end{eqnarray}
%------------------------------------------------------------------------   

\vspace{2mm}
\noindent
{\sc Definition.~}\\
A word $w$ is called a {\sc Lyndon} word if it is smaller than any of its 
suffixes.

\vspace*{3mm} \noindent
The operator $<$ between letters accounts for {\sf lexicographic
order}. This property offers an easy way to construct the set of {\sc
Lyndon} words for a given index set. 

One considers a commutative ring with unit, 
$K$. A {\sf non-commutative polynomial} on ${\mathfrak A}$ over $K$ is a
linear combination of words on ${\mathfrak A}$ with coefficients in $K$,
%------------------------------------------------------------------------
\begin{eqnarray}
\label{eqW2}
P = \sum_{w \epsilon {\mathfrak A}^*} (P,w) w. 
\end{eqnarray}
%------------------------------------------------------------------------
The set of all polynomials is denoted by $K\langle {\mathfrak A}\rangle$.
The algebra $K\langle {\mathfrak A}\rangle$ is the free associative 
$K$--algebra generated by ${\mathfrak A}$.

\vspace{2mm} \noindent
{\sc Theorem}({\sc Radford}~\cite{RADF}).\\
The shuffle algebra $K\langle{\mathfrak A}\rangle$ is freely generated by 
the {\sc Lyndon} words.

\vspace{3mm} \noindent
A direct conclusion is that the number of {\sc Lyndon} words counts the 
number of basis elements, which are the algebraically independent harmonic 
sums in our case. In the counting relations for the number of basic sums 
given below {\sc M\"{o}bius}' function $\mu(n)$~\cite{MOB}
with $n~\epsilon~{\bf N}$ emerges, which is defined by
%------------------------------------------------------------------------
\begin{eqnarray}
\label{eqMOB}
\mu(n) = \left\{\begin{array}{cl}
1 & n=1 \\
0 & n~~{\rm is~divided~by~the~square~of~a~prime}\\
(-1)^s & n~~{\rm is~the~product~of~{\it s}~different~primes.}
\end{array} \right.
\end{eqnarray}
%------------------------------------------------------------------------
The number of {\sc Lyndon} words of length $n$ over an alphabet of length $q$
is given by
%------------------------------------------------------------------------
\begin{eqnarray}
\label{eqWIT1}
l_n(q) = \frac{1}{n} \sum_{d|n} \mu(d) q^{n/d}~.
\end{eqnarray}
%------------------------------------------------------------------------
We will call this relation the first {\sc Witt} 
formula~\cite{WITT}.~\footnote{Usually it is called {\sf the} {\sc Witt} formula 
in the literature,~cf. e.g.~\cite{REUT}.} 
As mentioned in \cite{WITT} this relation also counts the number of prime polynomials
$\sum_{k=0}^n a_k x^{n-k}$ in the {\sc Galois} field of $q$ elements.
This relation has been known to {\sc Gauss} already,~\cite{LN}. 
As we would like to count the number of basic sums for all sums of a given index set
{\sf individually} this relation cannot be used, but we have to count the
number of
{\sc Lyndon} words belonging to this set. The respective
relation has been given in
the same paper as the second {\sc Witt} formula, 
%------------------------------------------------------------------------
\begin{eqnarray}
\label{eqWIT2}  
l_n(n_1, \ldots, n_q) = \frac{1}{n} \sum_{d|n_i} \mu(d)
\frac{\left(\frac{n}{d}\right)!}
{\left(\frac{n_1}{d}\right)! \ldots \left(\frac{n_q}{d}\right)!},~~~~~~~n 
= \sum_{k=1}^q n_k~.
\end{eqnarray}
%------------------------------------------------------------------------
One may derive the numbers $l_n(q)$ and $l_n(n_1, \ldots, n_q)$ using the
generating functionals \cite{WITT}~:
%------------------------------------------------------------------------
\begin{eqnarray}
\label{eqWIT1GF}
\frac{1}{1-q x} = \prod_{n=1}^\infty \left(\frac{1}{1-x^n}\right)^{l_n(q)}
\end{eqnarray}
%------------------------------------------------------------------------
and
%------------------------------------------------------------------------
\begin{eqnarray}
\label{eqWIT2GF} 
\frac{1}{1-x_1 - \ldots - x_{n_q}} =  \prod_{n = 1}^\infty
\left(\frac{1}{1-\sum_{k=1}^q x_k^{d_k}}\right)^{l_n(n_1, \ldots, n_q)}~.
\end{eqnarray}
%------------------------------------------------------------------------
Counting relations for multiple Zeta--values were also given in the literature.
Multiple Zeta--values are simpler objects than multiple harmonic sums and,
consequently, they obey more relations. Besides the shuffle--relations
so called stuffle and duality relations exist,~see e.g.~\cite{LIS,WALDS}.
{\sc Hoffman}~\cite{HOF} derived the {\sc Gauss--Witt} relation 
%------------------------------------------------------------------------
\begin{eqnarray}
\label{eqHOF}
N(w) = \frac{1}{w} \sum_{d|w} \mu\left(\frac{w}{d}\right) 2^d
\end{eqnarray}
%------------------------------------------------------------------------
for the number of basic multiple Zeta--values of weight $w$ 
for $\forall n_i > 0$. One
verifies that this is 
actually an upper bound and {\sc Broadhurst} and {\sc
Kreimer}~\cite{BRKR} conjectured that\footnote{See also {\sc Zagier}'s
conjecture~[16b]. {\sc Zagier} gave numerical checks up to weight $w=12$.} 
(\ref{eqHOF}) can be sharpened to
%------------------------------------------------------------------------
\begin{eqnarray}
\label{eqBRKR}  
N(w) = \frac{1}{w} \sum_{d|w} \mu\left(\frac{w}{d}\right) P_d~,
\end{eqnarray}
%------------------------------------------------------------------------
with $P_d$ the {\sc Perrin} numbers
%------------------------------------------------------------------------
\begin{eqnarray}
\label{eqPERR}
P_1 = 0, P_2 = 2, P_3 = 3, P_n = P_{n-2} + P_{n-3},~~~~n \geq 3~,
\end{eqnarray}
%------------------------------------------------------------------------
being checked up to $w =12$ in \cite{PET1}.  

It is clear from (\ref{eqPERM}) and (\ref{eqWIT2}) that the fraction of the 
number of basic sums to all sums of a given index set ${a_1, \ldots, a_n}$ is 
$1/n$ if the greatest common divisor (g.c.d.) of $n_1, \ldots, n_q$ is 1,
since the sum  (\ref{eqWIT2}) consists of only one term. 
If only one prime $p$ aside of 1
divides the numbers $n_1, \ldots, n_q$ (see the cases in the tables
above), 
the fraction is obviously smaller than $1/n$ since $\mu(p) = -1$. 
For the case of a fully symmetric harmonic
sum with the index set $\{a,a, \ldots, a\}$ (\ref{eqWIT2}) yields
%------------------------------------------------------------------------
\begin{eqnarray}
\label{eqSYMMHS}
l_n({a,a, \ldots, a}) = l_n(n) = \sum_{d|n} \mu(d)
\frac{\left(\frac{n}{d}\right)!}{\left(\frac{n}{d}\right)!} 
= \sum_{d|n} \mu(d) = 0,~~~~n > 1
\end{eqnarray}
%------------------------------------------------------------------------
due to the well--known property of the {\sc M\"obius} function. In
counting {\sf basic} harmonic sums we will leave out the trivial case
of single harmonic sums therefore\footnote{Their analytic continuation to
complex values of $N$ is trivially being obtained by the $\psi(z)$
function and their derivatives.}. Let us now study some explicit examples.

\vspace{2mm}\noindent
{\bf Index-Sets out of one Letter}\\
The letters are {\sc Lyndon} words. Words which are products out of a 
single letter $a a \ldots a$ with more than one factor are no {\sc Lyndon}
words. This property corresponds to the fact that harmonic sums of 
depth $d \geq 1$ with all indices equal can always be decomposed into a 
(symmetric) polynomial out of single harmonic sums, 
cf.~(\ref{eqS2}--\ref{eqS6}).

\vspace{2mm}\noindent
{\bf Index-Sets out of two Letters}\\
The first letter in a {\sc Lyndon} word has 
always to be the smallest letter of the alphabet emerging in the sequence,
which can never be the last letter. Words consisting out of (several) factors 
$a$ and a single factor $b$ are {\sc Lyndon} words only if they obey the 
sequence
%------------------------------------------------------------------------
\begin{equation}
\label{eqLW1}
a a \ldots a b~.
\end{equation}
%------------------------------------------------------------------------
Obviously $a b$ is a {\sc Lyndon} word and 
%------------------------------------------------------------------------
\begin{equation}
\underbrace{a \ldots a}_{ \small 
k} b < \underbrace{a \ldots a}_{\small 
k-1} b~. 
\end{equation}
%------------------------------------------------------------------------
Any emergence in  $a$ or its power in a word right of $b$ would violate
the order such that the word is no {\sc Lyndon} word. (\ref{eqLW1}) 
confirms the finding that matrix $M_{d;1}$~(\ref{eqMD}) is of rank $d-1$, i.e.
it exists one independent sum in these cases always. For depth $d=4$ the 
number of {\sc Lyndon} words for the sequences of type $\{a,a,b,b\}$ is 
one since $abba$ and $abab$ are no {\sc Lyndon} words. The second 
word is a power of a single {\sc Lyndon} word, which is no {\sc Lyndon} 
word and $aabb$ is the only {\sc Lyndon} word for this set. Words of 
length $d=5$ out of two different letters correspond to the set 
$\{a,a,a,b,b\}$, to which the two {\sc Lyndon} words $aaabb$ and 
$aabab$ belong. In general the following inequality  
holds~\cite{REUT} for {\sc Lyndon} words $u_1$ and $u_2$
%------------------------------------------------------------------------
\begin{equation}
u_1 < u_1^{\kappa_1} u_2^{\kappa_2} < u_2,~~~\forall 
\kappa_i~\epsilon~{\bf N}~,
\end{equation}
%------------------------------------------------------------------------
where $aab < ab$. For $d=6$  two {\sc Lyndon} words are associated to
the set $\{a,a,a,a,b,b\}$ 
synonymously. For the set $\{a,a,a,b,b,b\}$ three {\sc Lyndon} words 
exist, $aaabbb, aababb$ and $aabbab$. Since $aab < abb$ the first two 
cases are evident and the latter word is a {\sc Lyndon} word because
$aabb < ab$.

%-------------------------------------------------------------------------
\begin{center}
\begin{tabular}[h]{|rrrrr|}
\hline\hline
$ab$ & $aab$ & $aaab$ & $aaaab$ & $aaaaab$ \\
\hline
   &     & $aabb$ & $aaabb$ & $aaaabb$ \\
   &     &      & $aabab$ & $aaabab$ \\
\hline
   &     &      &       & $aaabbb$ \\
   &     &      &       & $aababb$ \\
   &     &      &       & $aabbab$ \\                   
\hline\hline
\end{tabular}
\end{center}
%-------------------------------------------------------------------------

\vspace{2mm}\noindent
{\bf Index-Sets out of only Different Letters}\\
The number of associated {\sc Lyndon} words can easily be obtained for
this case. The {\sc Lyndon} words have to begin with the letter $a$.
For words of length $n$ the {\sc Lyndon} words are obtained writing $a$
to the left of the $(n-1)!$ permutations of the letters $b_1, \ldots,
b_{n-1}$, i.e. the number of {\sc Lyndon} words is $(n-1)!$ and therefore
the $1/n$th of all possible combinations.

\vspace{2mm}\noindent
{\bf Numbers of Basic Sums: Examples}\\
Let us calculate the number of basic sums for a given index set for a few
examples. We denote by $n_k(\{a_1,...,a_q\})$
%-------------------------------------------------------------------------
\begin{eqnarray}
n_k(\{a_1,...,a_q\}) = \frac{k!}{n(a_1)! \ldots n(a_q)!}~,
\end{eqnarray}
%-------------------------------------------------------------------------
where $n(a_l)$ is the number of occurrences of $a_l$ in the string
$a_1,a_2, \ldots, a_q$.
For the set $\{a,a,a,a,b,b\}$ one obtains
%-------------------------------------------------------------------------
\begin{eqnarray}
l_6(\{a,a,a,a,b,b\}) &=& \frac{1}{6}\left[\mu(1)
\frac{6!}{4! 2!} + \mu(2)
\frac{3!}{2! 1!}\right] 
= 2,\\
\frac{l_6(\{a,a,a,a,b,b\})}{n_6(\{a,a,a,a,b,b\})} &=& \frac{2}{15} <
\frac{1}{6}~.
\end{eqnarray}
%-------------------------------------------------------------------------
Similarly one obtains for $\{a,a,a,b,b,b\}$
%-------------------------------------------------------------------------
\begin{eqnarray}
l_6(\{a,a,a,b,b,b\}) &=& \frac{1}{6}\left[\mu(1)
\frac{6!}{3! 3!} + \mu(3) \frac{2!}{1! 1!}\right]
= 3,\\
\frac{l_6(\{a,a,a,b,b,b\})}{n_6(\{a,a,a,b,b,b\})} &=& \frac{3}{20} <
\frac{1}{6}~.
\end{eqnarray}
%-------------------------------------------------------------------------
and for $\{a,a,b,b,c,c\}$
%-------------------------------------------------------------------------
\begin{eqnarray}
l_6(\{a,a,b,b,c,c\}) &=& \frac{1}{6}\left[\mu(1)
\frac{6!}{2! 2! 2!} + \mu(2) 
\frac{3!}{1! 1! 1!}\right]
= 14,\\
\frac{l_6(\{a,a,b,b,c,c\})}{n_6(\{a,a,b,b,c,c\})} &=& \frac{7}{45} <
\frac{1}{6}~.
\end{eqnarray}
%-------------------------------------------------------------------------
Finally,
%-------------------------------------------------------------------------
\begin{eqnarray}
l_{12}(\{a,a,a,a,a,a,b,b,b,b,b,b\}) &=& \frac{1}{12}\left[\mu(1)
\frac{12!}{6! 6!} + \mu(2)
\frac{6!}{3! 3!} + \mu(3) \frac{4!}{2! 2!} + \mu(6) \frac{2!}{1! 
  1!}\right]\nonumber\\
&=& 75,\\
\frac{l_6(\{a,a,a,a,a,a,b,b,b,b,b,b\})}
{n_6(\{a,a,a,a,a,a,b,b,b,b,b,b\})} &=& \frac{25}{308} <
\frac{1}{12}~. 
\end{eqnarray}
%-------------------------------------------------------------------------
We conclude that the fraction of the number of basic sums in all sums of a
given index set is primarily determined by the depth of the sum.

In the following Table we compare the number of harmonic sums with the
number
of basic sums determined applying the algebraic relations.
%---------------------------------------------------------------------------
\begin{center}
\begin{tabular}[h]{||r||r|r|r|r|l||}
\hline \hline %
\multicolumn{1}{||c||}{\sf Weight}&
\multicolumn{1}{c|}{\sf \# Sums}&
\multicolumn{1}{c|}{\sf Cum. \# Sums}&
\multicolumn{1}{c|}{\sf \# Basic Sums}&
\multicolumn{1}{c|}{\sf Cum. \# Basic Sums} &
\multicolumn{1}{c||}{\sf Cum. Fraction  }\\
\hline\hline
1 &   2  &   2 &   0  &   0 & 0.0    \\
2 &   6  &   8 &   1  &   1 & 0.1250 \\
3 &  18  &  26 &   6  &   7 & 0.2692 \\
4 &  54  &  80 &  16  &  23 & 0.2875 \\
5 & 162  & 242 &  46  &  69 & 0.2851 \\
6 & 486  & 728 & 114  & 183 & 0.2513 \\
\hline \hline  
\end{tabular}
\renewcommand{\arraystretch}{1.0}
\end{center}
%-----------------------------------------------------------------------------
The number of sums for a given weight is $2 \cdot 3^{w-1}$, with the 
cumulative value $3^w - 1$. The numbers of basic sums as a function of the
weight $w$ are given as sums over the 2nd {\sc Witt} formula.
The corresponding sequence seems not yet to be contained in {\sc Sloane}'s
on-line encyclopedia of integer sequences~\cite{SLOA}.
%%%%%%%%%%%%%%%%%%%%%%%%%%%%%%%%%%%%%%%%%%%%%%%%%%%%%%%%%%%%%%%%%%%%%%%%
\section{Conclusions}
%%%%%%%%%%%%%%%%%%%%%%%%%%%%%%%%%%%%%%%%%%%%%%%%%%%%%%%%%%%%%%%%%%%%%%%%
%

\vspace{1mm}
\noindent
The product of finite alternating or non--alternating harmonic sums is given
by 
the shuffle product of harmonic sums and polynomials of harmonic sums of lower 
depth. These representations imply algebraic relations between the harmonic 
sums. If one considers all harmonic sums associated to an index set
$\{a_1, \ldots, a_k\}$ one may express these sums by a number of basic
sums. It turns out that this number is given by the 2nd {\sc Witt}
formula which counts the
number of {\sc Lyndon} words corresponding to the respective index set.
The set of these {\sc Lyndon} words generates in this sense all harmonic
sums of this class {\sf freely}. By solving the corresponding linear
equations we
derived the explicit representation of all harmonic sums up to depth $d=6$
without specifying the indices numerically and gave all expression which
are structurally needed to express the sums up to weight $w=6$. The
counting relations for the basis of the finite harmonic sums were given 
up to depth $d=10$. The relations derived hold likewise for other 
mathematical objects obeying the same multiplication relation or a simpler
one, which is being contained, as that for harmonic polylogarithms. This
is due to the fact that the relations derived depend on the index set
and the multiplication relation but on no further properties of the
objects considered.  
  
The ratio of the number of basic sums for a given index set and 
the number of all sums is mainly determined by the depth $d$ rather
than the weight of the respective sums, due to the pre-factor $1/d$ in the 
{\sc Witt} formula. Modifications occur due to common non-trivial divisors   
of the numbers of individual indices in the set being considered. Up to
$d=10$ we showed that the fraction of basic sums is always $\leq 1/d$
compared to all sums. The use of these algebraic relations leads to a
considerable reduction in the set of functions needed to express the
results of higher order calculations in massless QED and QCD and related
subjects. For practical applications such as the description of the QCD 
scaling violation of the structure functions in deeply inelastic
scattering the harmonic sums  occurring in the {\sc Mellin} $N$ space
calculation have to be translated to $x$--space by the inverse {\sc
Mellin} transform. For this reason the respective harmonic sums have to be 
analytically continued in the argument $N$ to complex values, which
requires a high effort using numerical procedures. It is therefore
recommended to use as many as possible relations between the $N$ space 
objects before to perform the last step only for a reduced set.

\vspace*{4mm}
\noindent
{\bf Acknowledgment.}~~For useful conversations I would like to thank
H. Gangl, S. Moch, J. Vermaseren, S. Weinzierl, P. Will, and H. Gangl for
a careful reading of the manuscript.

%%%%%%%%%%%%%%%%%%%%%%%%%%%%%%%%%%%%%%%%%%%%%%%%%%%%%%%%%%%%%%%%%%%%%%%%
\newpage
\section{Appendix~A: Number of Harmonic Sums and Basic Sums for
Individual Index-Pattern of Depth 7 to 10} 
%%%%%%%%%%%%%%%%%%%%%%%%%%%%%%%%%%%%%%%%%%%%%%%%%%%%%%%%%%%%%%%%%%%%%%%% 
%

{\footnotesize
\begin{center}
\renewcommand{\arraystretch}{1.3}
\begin{tabular}[h]{||l||r|r|c||}
\hline \hline 
\multicolumn{1}{||c||}{\sf Index Set}&
\multicolumn{1}{c|}{\sf Number of Sums}&
\multicolumn{1}{c|}{\sf Basic Sums}&
\multicolumn{1}{c||}{\sf Frac. of Sums} \\
\hline \hline
$\{a,a,a,a,a,a,a\}$ &    1 &   0 &   0 \\
\hline
$\{a,a,a,a,a,a,b\}$ &    7 &   1 & 1/7 \\
$\{a,a,a,a,a,b,b\}$ &   21 &   3 & 1/7 \\
$\{a,a,a,a,b,b,b\}$ &   35 &   5 & 1/7 \\
\hline
$\{a,a,a,a,a,b,c\}$ &   42 &   6 & 1/7 \\
$\{a,a,a,a,b,b,c\}$ &  105 &  15 & 1/7 \\
$\{a,a,a,b,b,b,c\}$ &  140 &  20 & 1/7 \\
$\{a,a,a,b,b,c,c\}$ &  210 &  30 & 1/7 \\
\hline
$\{a,a,a,a,b,c,d\}$ &  210 &  30 & 1/7 \\
$\{a,a,a,b,b,c,d\}$ &  420 &  60 & 1/7 \\
$\{a,a,b,b,c,c,d\}$ &  630 &  90 & 1/7 \\
\hline
$\{a,a,a,b,c,d,e\}$ &  840 & 120 & 1/7 \\
$\{a,a,b,b,c,d,e\}$ & 1260 & 180 & 1/7 \\
\hline
$\{a,a,b,c,d,e,f\}$ & 2550 & 360 & 1/7 \\
\hline
$\{a,b,c,d,e,f,g\}$ & 5040 & 720 & 1/7 \\
\hline \hline                   
\end{tabular}

\renewcommand{\arraystretch}{1.0}
\end{center}
}

{\footnotesize
\begin{center}
\renewcommand{\arraystretch}{1.3}
\begin{tabular}[h]{||l||r|r|c||}
\hline \hline 
\multicolumn{1}{||c||}{\sf Index Set}&
\multicolumn{1}{c|}{\sf Number of Sums}&
\multicolumn{1}{c|}{\sf Basic Sums}&
\multicolumn{1}{c||}{\sf Frac. of Sums} \\
\hline \hline
$\{a,a,a,a,a,a,a,a\}$ &     1 &    0&     0 \\
\hline
$\{a,a,a,a,a,a,a,b\}$ &     8 &    1&   1/8 \\
$\{a,a,a,a,a,a,b,b\}$ &    28 &    3&  3/28 \\
$\{a,a,a,a,a,b,b,b\}$ &    56 &    7&   1/7 \\
$\{a,a,a,a,b,b,b,b\}$ &    70 &    8&   4/35\\
\hline
$\{a,a,a,a,a,a,b,c\}$ &    56 &    7&   1/8 \\
$\{a,a,a,a,a,b,b,c\}$ &   168 &   21&   1/8 \\
$\{a,a,a,a,b,b,b,c\}$ &   280 &   35&   1/8 \\
$\{a,a,a,a,b,b,c,c\}$ &   420 &   51&   17/140 \\
$\{a,a,a,b,b,b,c,c\}$ &   560 &   70&   1/8 \\
\hline
$\{a,a,a,a,a,b,c,d\}$ &   336 &   42&   1/8 \\
$\{a,a,a,a,b,b,c,d\}$ &   840 &  105&   1/8 \\
$\{a,a,a,b,b,b,c,d\}$ &  1120 &  140&   1/8 \\
$\{a,a,a,b,b,c,c,d\}$ &  1680 &  210&   1/8 \\
$\{a,a,b,b,c,c,d,d\}$ &  2520 &  312&   13/105 \\
\hline
$\{a,a,a,a,b,c,d,e\}$ &  1680 &  210&   1/8 \\ 
$\{a,a,a,b,b,c,d,e\}$ &  3360 &  420&   1/8 \\
$\{a,a,b,b,c,c,d,e\}$ &  5040 &  630&   1/8 \\
\hline
$\{a,a,a,b,c,d,e,f\}$ &  6720 &  840&   1/8 \\
$\{a,a,b,b,c,d,e,f\}$ & 10080 & 1260&   1/8 \\
\hline
$\{a,a,b,c,d,e,f,g\}$ & 20160 & 2520&   1/8 \\
\hline
$\{a,b,c,d,e,f,g,h\}$ & 40320 & 5040&   1/8 \\
\hline\hline
\end{tabular}

\renewcommand{\arraystretch}{1.0}
\end{center}
}

{\footnotesize
\begin{center}
\renewcommand{\arraystretch}{1.3}
\begin{tabular}[h]{||l||r|r|c||}
\hline \hline 
\multicolumn{1}{||c||}{\sf Index Set}&
\multicolumn{1}{c|}{\sf Number of Sums}&
\multicolumn{1}{c|}{\sf Basic Sums}&
\multicolumn{1}{c||}{\sf Frac. of Sums} \\
\hline\hline
$\{a,a,a,a,a,a,a,a,a\}$ &      1 &     0  &    0 \\
\hline
$\{a,a,a,a,a,a,a,a,b\}$ &      9 &     1  &  1/9 \\
$\{a,a,a,a,a,a,a,b,b\}$ &     36 &     4  &  1/9 \\
$\{a,a,a,a,a,a,a,b,b\}$ &     72 &     8  &  1/9 \\
$\{a,a,a,a,a,a,b,b,b\}$ &     84 &     9  & 3/28 \\
$\{a,a,a,a,a,b,b,b,b\}$ &    126 &    14  &  1/9 \\
\hline
$\{a,a,a,a,a,a,b,b,c\}$ &    252 &    28  &  1/9 \\
$\{a,a,a,a,a,b,b,b,c\}$ &    504 &    56  &  1/9 \\
$\{a,a,a,a,a,b,b,c,c\}$ &    756 &    84  &  1/9 \\
$\{a,a,a,a,b,b,b,b,c\}$ &    630 &    70  &  1/9 \\
$\{a,a,a,a,b,b,b,c,c\}$ &   1260 &   140  &  1/9 \\
$\{a,a,a,b,b,b,c,c,c\}$ &   1680 &   186  &  31/280 \\
\hline
$\{a,a,a,a,a,a,b,c,d\}$ &    504 &    56  &  1/9 \\
$\{a,a,a,a,a,b,b,c,d\}$ &   1512 &   168  &  1/9 \\
$\{a,a,a,a,b,b,b,c,d\}$ &   2550 &   280  &  1/9 \\
$\{a,a,a,a,b,b,c,c,d\}$ &   3780 &   420  &  1/9 \\
$\{a,a,a,b,b,b,c,c,d\}$ &   5040 &   560  &  1/9 \\
$\{a,a,a,b,b,c,c,d,d\}$ &   7560 &   840  &  1/9 \\
\hline
$\{a,a,a,a,a,b,c,d,e\}$ &   3024 &   336  &  1/9 \\
$\{a,a,a,a,b,b,c,d,e\}$ &   7560 &   840  &  1/9 \\
$\{a,a,a,b,b,b,c,d,e\}$ &  10080 &  1120  &  1/9 \\
$\{a,a,a,b,b,c,c,d,e\}$ &  15120 &  1680  &  1/9 \\
$\{a,a,b,b,c,c,d,d,e\}$ &  22680 &  2520  &  1/9 \\
\hline
$\{a,a,a,a,b,c,d,e,f\}$ &  15120 &  1680  &  1/9 \\
$\{a,a,a,b,b,c,d,e,f\}$ &  30240 &  3360  &  1/9 \\
$\{a,a,b,b,c,c,d,e,f\}$ &  45360 &  5040  &  1/9 \\
\hline
$\{a,a,a,b,c,d,e,f,g\}$ &  60480 &  6720  &  1/9 \\
$\{a,a,b,b,c,d,e,f,g\}$ &  90720 & 10080  &  1/9 \\
\hline
$\{a,a,b,c,d,e,f,g,h\}$ &  90720 & 10080  &  1/9 \\
\hline
$\{a,b,c,d,e,f,g,h,i\}$ & 362880 & 40320  &  1/9 \\
\hline\hline
\end{tabular}

\renewcommand{\arraystretch}{1.0}
\end{center}
}

{\footnotesize
\begin{center}
\renewcommand{\arraystretch}{1.3}
\begin{tabular}[h]{||l||r|r|c||}
\hline \hline 
\multicolumn{1}{||c||}{\sf Index Set}&
\multicolumn{1}{c|}{\sf Number of Sums}&
\multicolumn{1}{c|}{\sf Basic Sums}&
\multicolumn{1}{c||}{\sf Frac. of Sums} \\
\hline\hline
$\{a,a,a,a,a,a,a,a,a,a\}$ &       1 &      0 &  0  \\
\hline
$\{a,a,a,a,a,a,a,a,a,b\}$ &      10 &      1 &  1/10\\
$\{a,a,a,a,a,a,a,a,b,b\}$ &      45 &      4 &  4/45\\
$\{a,a,a,a,a,a,a,b,b,b\}$ &     120 &     12 &  1/10\\
$\{a,a,a,a,a,a,b,b,b,b\}$ &     210 &     20 &  2/21\\
$\{a,a,a,a,a,b,b,b,b,b\}$ &     252 &     25 & 25/252\\
\hline
$\{a,a,a,a,a,a,a,a,b,c\}$ &      90 &      9 & 1/10\\
$\{a,a,a,a,a,a,a,b,b,c\}$ &     360 &     36 & 1/10\\
$\{a,a,a,a,a,a,b,b,b,c\}$ &     840 &     84 & 1/10\\
$\{a,a,a,a,a,a,b,b,c,c\}$ &    1260 &    124 & 31/315\\
$\{a,a,a,a,a,b,b,b,b,c\}$ &    1260 &    126 & 1/10\\
$\{a,a,a,a,a,b,b,b,c,c\}$ &    2520 &    252 & 1/10\\
$\{a,a,a,a,b,b,b,b,c,c\}$ &    3150 &    312 & 52/525\\
$\{a,a,a,a,b,b,b,c,c,c\}$ &    4200 &    420 & 1/10\\
\hline
$\{a,a,a,a,a,a,a,b,c,d\}$ &     720 &     72 & 1/10\\
$\{a,a,a,a,a,a,b,b,c,d\}$ &    2520 &    252 & 1/10\\
$\{a,a,a,a,a,b,b,b,c,d\}$ &    5040 &    504 & 1/10\\
$\{a,a,a,a,a,b,b,c,c,d\}$ &    7560 &    756 & 1/10\\
$\{a,a,a,a,b,b,b,b,c,d\}$ &    6300 &    630 & 1/10\\
$\{a,a,a,a,b,b,b,c,c,d\}$ &   12600 &   1260 & 1/10\\
$\{a,a,a,a,b,b,c,c,d,d\}$ &   18900 &   1884 & 157/1575\\
$\{a,a,a,b,b,b,c,c,c,d\}$ &   16800 &   1680 & 1/10 \\
$\{a,a,a,b,b,b,c,c,d,d\}$ &   25200 &   2520 & 1/10 \\
\hline
$\{a,a,a,a,a,a,b,c,d,e\}$ &    5040 &    504 & 1/10\\
$\{a,a,a,a,a,b,b,c,d,e\}$ &   15120 &   1520 & 1/10 \\
$\{a,a,a,a,b,b,b,c,d,e\}$ &   25200 &   2520 & 1/10 \\
$\{a,a,a,a,b,b,c,c,d,e\}$ &   37800 &   3780 & 1/10 \\
$\{a,a,a,b,b,b,c,c,d,e\}$ &   50400 &   5040 & 1/10 \\
$\{a,a,a,b,b,c,c,d,d,e\}$ &   75600 &   7560 & 1/10 \\
$\{a,a,b,b,c,c,d,d,e,e\}$ &  113400 &  11328 & 472/4725 \\
\hline
$\{a,a,a,a,a,b,c,d,e,f\}$ &   30240 &   3024 & 1/10\\
$\{a,a,a,b,b,b,c,d,e,f\}$ &   75600 &   7560 & 1/10 \\
$\{a,a,a,a,b,b,c,d,e,f\}$ &  100800 &  10080 & 1/10 \\
$\{a,a,a,b,b,c,c,d,e,f\}$ &  151200 &  15120 & 1/10 \\
$\{a,a,b,b,c,c,d,d,e,f\}$ &  226900 &  22680 & 1/10 \\
\hline
$\{a,a,a,a,b,c,d,e,f,g\}$ &  151200 &  15120 & 1/10 \\
$\{a,a,a,b,b,c,d,e,f,g\}$ &  302400 &  30240 & 1/10 \\
$\{a,a,b,b,c,c,d,e,f,g\}$ &  453600 &  45360 & 1/10 \\ 
\hline
$\{a,a,a,b,c,d,e,f,g,h\}$ &  302400 &  30240 & 1/10 \\
$\{a,a,b,b,c,d,e,f,g,h\}$ &  907200 &  90720 & 1/10 \\
\hline
$\{a,a,b,c,d,e,f,g,h,i\}$ & 1814400 & 181440 & 1/10 \\
\hline
$\{a,b,c,d,e,f,g,h,i,j\}$ & 3628800 & 362880 & 1/10 \\
\hline\hline
\end{tabular}

\renewcommand{\arraystretch}{1.0}
\end{center}
}

%%%%%%%%%%%%%%%%%%%%%%%%%%%%%%%%%%%%%%%%%%%%%%%%%%%%%%%%%%%%%%%%%%%%%%%%
\newpage
\section{Appendix~B: Overview on all Harmonic Sums up to Depth and
Weight~6}
%%%%%%%%%%%%%%%%%%%%%%%%%%%%%%%%%%%%%%%%%%%%%%%%%%%%%%%%%%%%%%%%%%%%%%%% 
%
{\footnotesize
\begin{multicols}{2}
\raggedcolumns
%
%\vspace{1mm}\nonumber
\begin{center}
\renewcommand{\arraystretch}{1.3}
\begin{tabular}[h]{||l|c|c|c||}
\hline \hline
\multicolumn{1}{||c|}{\sf Index Set}&
\multicolumn{1}{c|}{\sf Number}&  
\multicolumn{1}{c|}{\sf  Weight}&
\multicolumn{1}{c||}{\sf  Relations}
\\
\hline\hline
$\{-1\}$               &   1  &  1 & 1 \\ 
$\{1\}$                &   1  &  1 & 1 \\
\hline
$\{-2\}$                &   1  &  2 & 1 \\
$\{2\}$                 &   1  &  2 & 1 \\
$\{-1,-1\}$             &   1  &  2 & 1 \\
$\{-1,1\}$              &   2  &  2 & 1 \\
$\{1,1\}$               &   1  &  2 & 1 \\
\hline
$\{-3\}$                &   1  &  3 & 1 \\
$\{ 3\}$                &   1  &  3 & 1 \\
$\{-2,-1\}$             &   2  &  3 & 1 \\
$\{-2,1\}$              &   2  &  3 & 1 \\
$\{2,-1\}$              &   2  &  3 & 1 \\
$\{2,1\}$               &   2  &  3 & 1 \\
$\{-1,-1,-1\}$          &   1  &  3 & 1 \\
$\{-1,-1,1\}$           &   3  &  3 & 2 \\
$\{-1,1,1\}$            &   3  &  3 & 2 \\
$\{1,1,1\}$             &   1  &  3 & 1 \\
\hline
$\{-4\}$                &   1  &  4 & 1 \\   
$\{ 4\}$                &   1  &  4 & 1 \\
$\{-2,-2\}$             &   1  &  4 & 1 \\
$\{-2,2\}$              &   2  &  4 & 1 \\
$\{2,2\}$               &   1  &  4 & 1 \\
$\{-3,-1\}$             &   2  &  4 & 1 \\
$\{-3,1\}$              &   2  &  4 & 1 \\
$\{3,-1\}$              &   2  &  4 & 1 \\
$\{3,1\}$               &   2  &  4 & 1 \\
$\{-2,-1,-1\}$          &   3  &  4 & 2 \\
$\{-2,-1,1\}$           &   6  &  4 & 4 \\ 
$\{-2,1,1\}$            &   3  &  4 & 2 \\  
$\{2,-1,-1\}$           &   3  &  4 & 2 \\
$\{2,-1,1\}$            &   6  &  4 & 4 \\
$\{2,1,1\}$             &   3  &  4 & 2 \\
$\{-1,-1,-1,-1\}$       &   1  &  4  & 1\\
$\{-1,-1,-1,1\}$        &   4  &  4  & 3\\
$\{-1,-1,1,1\}$         &   6  &  4  & 5\\
$\{-1,1,1,1\}$          &   4  &  4  & 3\\
$\{1,1,1,1\}$           &   1  &  4  & 1\\
\hline \hline
\end{tabular}
\renewcommand{\arraystretch}{1.0}
\end{center}
%\newpage
%
%
%\vspace*{-4mm}
\noindent
\begin{center}
\renewcommand{\arraystretch}{1.3}
\begin{tabular}[h]{||l|c|c|c||}
\hline \hline
\multicolumn{1}{||c|}{\sf Index Set}&
\multicolumn{1}{c|}{\sf Number}&  
\multicolumn{1}{c||}{\sf  Weight}&
\multicolumn{1}{c||}{\sf  Relations}
\\
\hline\hline
$\{-5\}$                &   1  &  5  & 1\\
$\{ 5\}$                &   1  &  5  & 1\\
$\{-4,-1\}$             &   2  &  5  & 1\\
$\{-4,1\}$              &   2  &  5  & 1\\
$\{4,-1\}$              &   2  &  5  & 1\\
$\{4,1\}$               &   2  &  5  & 1\\
$\{-3,-2\}$             &   2  &  5  & 1\\
$\{-3,2\}$              &   2  &  5  & 1\\
$\{3,-2\}$              &   2  &  5  & 1\\
$\{3,2\}$               &   2  &  5  & 1\\
$\{-3,-1,-1\}$          &   3  &  5  & 2\\
$\{-3,-1,1\}$           &   6  &  5  & 4\\
$\{-3,1,1\}$            &   3  &  5  & 2\\
$\{3,-1,-1\}$           &   3  &  5  & 2\\
$\{3,-1,1\}$            &   6  &  5  & 4\\
$\{3,1,1\}$             &   3  &  5  & 2\\
$\{-2,-2,-1\}$          &   3  &  5  & 2\\
$\{-2,-2,1\}$           &   3  &  5  & 2\\
$\{-2,2,-1\}$           &   6  &  5  & 4\\
$\{-2,2,1\}$            &   6  &  5  & 4\\
$\{2,2,-1\}$            &   3  &  5  & 2\\
$\{2,2,1\}$             &   3  &  5  & 2\\
$\{-2,-1,-1,-1\}$       &   4  &  5  & 3\\
$\{-2,-1,-1,1\}$        &  12  &  5  & 9\\
$\{-2,-1,1,1\}$         &  12  &  5  & 9 \\
$\{-2,1,1,1\}$          &   4  &  5  & 3\\
$\{2,-1,-1,-1\}$        &   4  &  5  & 3\\
$\{2,-1,-1,1\}$         &  12  &  5  & 9\\
$\{2,-1,1,1\}$          &  12  &  5  & 9\\
$\{2,1,1,1\}$           &   4  &  5  & 3\\
$\{-1,-1,-1,-1,-1\}$    &   1  &  5 &  1\\
$\{-1,-1,-1,-1, 1\}$    &   5  &  5 &  4\\
$\{-1,-1,-1, 1, 1\}$    &  10  &  5 &  8\\
$\{-1,-1, 1, 1, 1\}$    &  10  &  5 &  8\\
$\{-1, 1, 1, 1, 1\}$    &   5  &  5 &  4\\
$\{ 1, 1, 1, 1, 1\}$    &   1  &  5 &  1\\
\hline \hline
\end{tabular}
\renewcommand{\arraystretch}{1.0}
\end{center}
%\vspace{1mm}\nonumber
\begin{center}
\renewcommand{\arraystretch}{1.3}
\begin{tabular}[t]{||l|c|c|c||}
\hline \hline
\multicolumn{1}{||c|}{\sf Index Set}&
\multicolumn{1}{c|}{\sf Number}&  
\multicolumn{1}{c||}{\sf  Weight}&
\multicolumn{1}{c||}{\sf  Relations}
\\
\hline\hline
$\{-6\}$                &   1  &  6 &  1\\
$\{ 6\}$                &   1  &  6 &  1\\
$\{-5,-1\}$             &   2  &  6 &  1\\
$\{-5,1\}$              &   2  &  6 &  1\\
$\{5,-1\}$              &   2  &  6 &  1\\
$\{5,1\}$               &   2  &  6 &  1\\
$\{-4,-2\}$             &   2  &  6 &  1\\
$\{-4,2\}$              &   2  &  6 &  1\\ 
$\{4,-2\}$              &   2  &  6 &  1\\
$\{4,2\}$               &   2  &  6 &  1\\
$\{-3,-3\}$             &   1  &  6 &  1\\
$\{-3,3\}$              &   2  &  6 &  1\\
$\{3,3\}$               &   1  &  6 &  1\\
$\{-2,-2,-2\}$          &   1  &  6 &  1\\
$\{-2,-2, 2\}$          &   3  &  6 &  2\\
$\{-2, 2, 2\}$          &   3  &  6 &  2\\
$\{ 2, 2, 2\}$          &   1  &  6 &  1\\
$\{-4,-1,-1\}$          &   3  &  6 &  2\\
$\{-4,-1,1\}$           &   6  &  6 &  4\\
$\{-4,1,1\}$            &   3  &  6 &  2\\
$\{4,-1,-1\}$           &   3  &  6 &  2\\
$\{4,-1,1\}$            &   6  &  6 &  4\\
$\{4,1,1\}$             &   3  &  6 &  2\\
$\{-3,-2,-1\}$          &   6  &  6 &  4\\
$\{-3,-2,1\}$           &   6  &  6 &  4\\
$\{-3,2,-1\}$           &   6  &  6 &  4\\
$\{-3,2,1\}$            &   6  &  6 &  4\\
$\{3,-2,-1\}$           &   6  &  6 & 4\\
$\{3,-2,1\}$            &   6  &  6 & 4\\
$\{3,2,-1\}$            &   6  &  6 & 4\\
$\{3,2,1\}$             &   6  &  6 & 4\\
$\{-3,-1,-1,-1\}$       &   4  &  6 & 3\\
$\{-3,-1,-1, 1\}$       &  12  &  6 & 9\\
$\{-3,-1, 1, 1\}$       &  12  &  6 & 9\\
$\{-3, 1, 1, 1\}$       &   4  &  6 & 3\\
$\{ 3,-1,-1,-1\}$       &   4  &  6 & 3\\
$\{ 3,-1,-1, 1\}$       &  12  &  6 & 9\\
$\{ 3,-1, 1, 1\}$       &  12  &  6 & 9\\
$\{ 3, 1, 1, 1\}$       &   4  &  6 & 3\\
\hline \hline
\end{tabular}
\renewcommand{\arraystretch}{1.0}
\end{center}
%
%\vspace{-5mm}
\noindent
\begin{center}
\renewcommand{\arraystretch}{1.3}
\begin{tabular}[t]{||l|c|c|c||}
\hline \hline
\multicolumn{1}{||c|}{\sf Index Set}&
\multicolumn{1}{c|}{\sf Number}&  
\multicolumn{1}{c||}{\sf  Weight}&
\multicolumn{1}{c||}{\sf  Relations}
\\
\hline\hline
$\{-2,-2,-1,-1\}$       &   6  &  6 & 5\\
$\{-2,-2,-1, 1\}$       &  12  &  6 & 9\\
$\{-2,-2, 1, 1\}$       &   6  &  6 & 5\\
$\{-2, 2,-1,-1\}$       &  12  &  6 & 9\\
$\{-2, 2,-1, 1\}$       &  24  &  6 & 18\\
$\{-2, 2, 1, 1\}$       &  12  &  6 & 9\\
$\{ 2, 2,-1,-1\}$       &   6  &  6 & 5\\
$\{ 2, 2,-1, 1\}$       &  12  &  6 & 9\\
$\{ 2, 2, 1, 1\}$       &   6  &  6 & 5\\
$\{-2,-1,-1,-1,-1\}$    &   5  &  6 & 4\\
$\{-2,-1,-1,-1, 1\}$    &  20  &  6 & 16\\
$\{-2,-1,-1, 1, 1\}$    &  30  &  6 & 24\\
$\{-2,-1, 1, 1, 1\}$    &  20  &  6 & 16 \\
$\{-2, 1, 1, 1, 1\}$    &   5  &  6 & 4\\
$\{ 2,-1,-1,-1,-1\}$    &   5  &  6 & 4\\
$\{ 2,-1,-1,-1, 1\}$    &  20  &  6 & 16\\
$\{ 2,-1,-1, 1, 1\}$    &  30  &  6 & 24\\
$\{ 2,-1, 1, 1, 1\}$    &  20  &  6 & 16\\
$\{ 2, 1, 1, 1, 1\}$    &   5  &  6 & 4\\
$\{-1,-1,-1,-1,-1,-1\}$ &   1  &  6 &  1\\
$\{-1,-1,-1,-1,-1, 1\}$ &   6  &  6 &  5\\
$\{-1,-1,-1,-1, 1, 1\}$ &  15  &  6 & 13\\
$\{-1,-1,-1, 1, 1, 1\}$ &  20  &  6 & 17\\
$\{-1,-1, 1, 1, 1, 1\}$ &  15  &  6 & 13\\
$\{-1, 1, 1, 1, 1, 1\}$ &   6  &  6 &  5\\
$\{ 1, 1, 1, 1, 1, 1\}$ &   1  &  6 &  1\\
\hline \hline
\end{tabular}
\renewcommand{\arraystretch}{1.0}
\end{center}
\end{multicols}
}
%%%%%%%%%%%%%%%%%%%%%%%%%%%%%%%%%%%%%%%%%%%%%%%%%%%%%%%%%%%%%%%%%%%%%%%%%
%%%%%%%%%%%%%%%%%%%%%%%%%%%%%%%%%%%%%%%%%%%%%%%%%%%%%%%%%%%%%%%%%%%%%%%%%

\newpage
%%%%%%%%%%%%%%%%%%%%%%%%%%%%%%%%%%%%%%%%%%%%%%%%%%%%%%%%%%%%%%%%%%%%%%%%%

%%%%%%%%%%%%%%%%%%%%%%%%%%%%%%%%%%%%%%%%%%%%%%%%%%%%%%%%%%%%%%%%%%%%%%
\end{document}